\documentclass[a4paper,12pt,oneside]{book}

\usepackage[utf8]{inputenc}
\usepackage[english]{babel}

\usepackage{amsmath}
\usepackage{amsthm}   
\usepackage{amsfonts}
\usepackage{amssymb}
\usepackage{mathrsfs} 
\usepackage{graphicx}
\usepackage{epigraph}
\usepackage{fancyhdr}
\usepackage{braket}
\usepackage{wrapfig}
\usepackage{hyperref}
\usepackage{bm}
\usepackage[font=small,format=hang,labelfont={sf,bf}]{caption}
\usepackage{mathptmx}
\usepackage{float}
\usepackage{subfigure}

\fancypagestyle{plain}{%
\fancyhead{} % leva l’intestazione
}

\def\rme{{\rm {e}}}
\def\rmi{{\rm {i}}}

\def\be{\begin{equation}}
\def\ee{\end{equation}}
\def\ba{\begin{eqnarray}}
\def\ea{\end{eqnarray}}

\newcommand{\he}{\mathcal{H}}
\newcommand{\te}{\mathcal{T}}
\newcommand{\E}{\mathcal{E}}
\addto\captionsitalian{}

  {\left\lbrace\begin{array}{@{}l@{}}}%
  {\end{array}\right.}

\newcommand{\citazioneA}[2]{\begin{flushright}% 
    #1 \\ \medskip% 
    --- #2%      
\end{flushright}} 
 
\newenvironment{mychapter}[1][]{\chapter*{#1\markboth{#1}{}\addcontentsline{toc}{chapter}{#1}}}{}

\title{----} 
\author{Biella Alberto}
\date{A.A. ----}

\begin{document}
\frontmatter
\thispagestyle{empty}
\vspace*{-1.2cm} 
{\bfseries
\begin{center}
\large
\textsc{UNIVERSIT\`{A} CATTOLICA DEL SACRO CUORE \mbox{SEDE DI BRESCIA}}\\[1mm]
\normalsize
Facolt\`{a} di Scienze Matematiche, Fisiche e Naturali\\
\vspace*{0.3 cm}
\normalsize
Corso di Laurea in Fisica\\
\vspace*{0.3 cm}
\begin{figure}[h]
\begin{center}
\scalebox{0.3}{\includegraphics{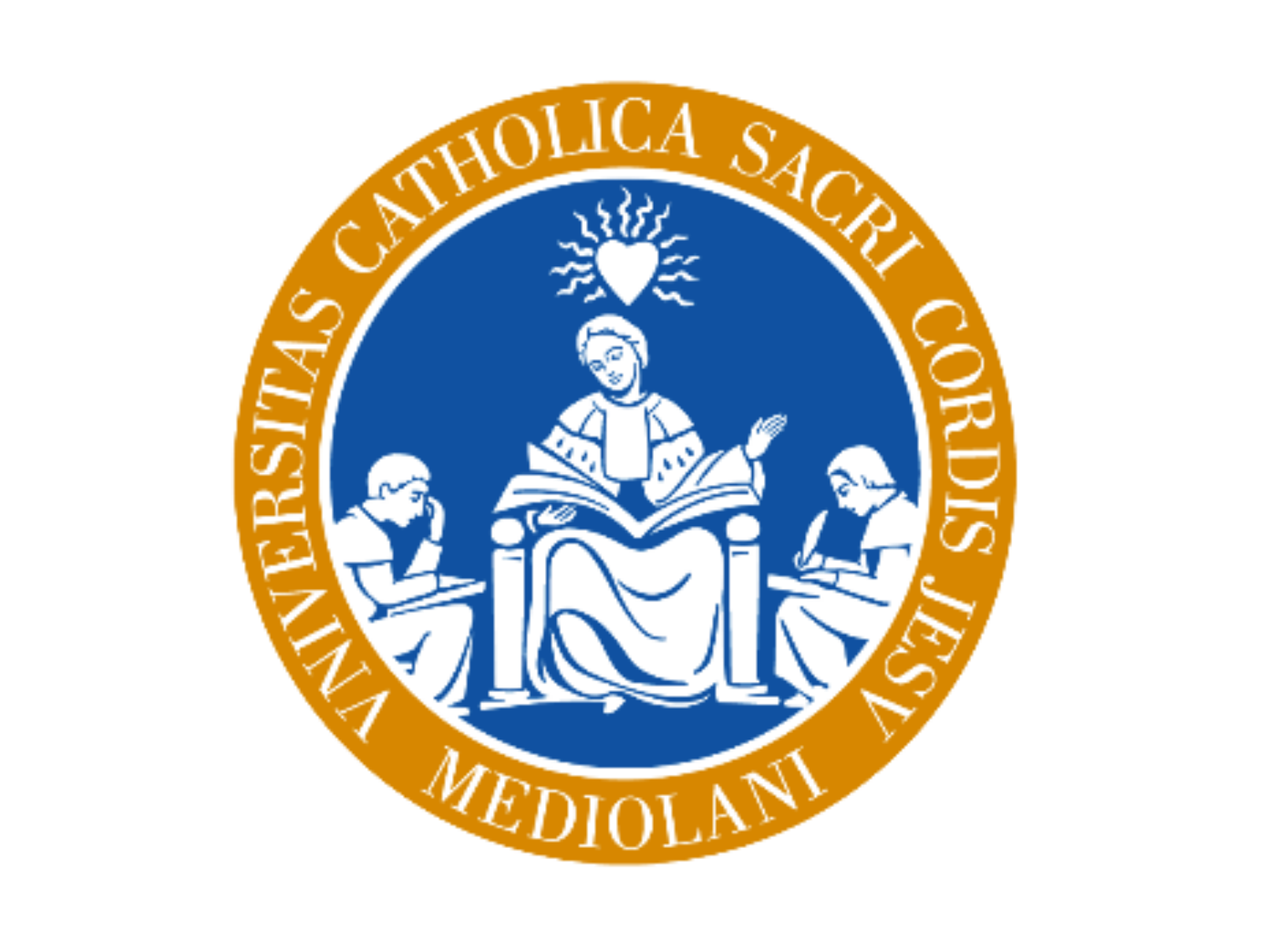}}
\end{center}
\end{figure}
  \vspace*{0.2 cm}
  Tesi di Laurea\\
  \vspace*{0.6 cm} \LARGE
\textbf{From Dicke to Anderson: \mbox{Interplay of Superradiance and Disorder}}\\
\vspace*{1.5 cm}
\normalsize  
\begin{tabular}{l c l}
 \textit{Relatore:}   & & \\
 Prof. F. Borgonovi& & \\[3mm]
 \textit{Correlatore:}   & & \\
 Dott. G. L. Celardo& & \\
 & & \\ 
 & \hspace{3.5cm} & \textit{Candidato:} \\
 & \hspace{3.5cm} & Alberto Biella  \\
  & \hspace{3.5cm} & \textit{Matricola n.} 3909760 \\
  & & \\
\end{tabular} 
\vspace*{0.8 cm}
\newline
Anno Accademico 2011-2012
\end{center} \clearpage}

\bigskip

\citazioneA{Quantum phenomena do not occur in a Hilbert space. \\They occur in a laboratory.}{A. Peres\\
Quantum theory: concepts and methods, 1995} 

\vspace{3cm}

\citazioneA{Localization [..], very few believed it at the time, \\and even fewer saw its importance, \\among those who failed to fully understand it \\at first was certainly its author. \\It has yet to receive adequate mathematical treatment, \\and one has to resort to the indignity of numerical simulations \\to settle even the simplest questions about it.}{P.W. Anderson\\ Nobel Lecture, 1977}

\vspace{3cm}

\citazioneA{Galileo chi si oppose al tuo genio\\
fu più vil del coyote nel canyon,\\
se la chiesa ti ha messo all'indice\\
beh che male c'è tu la metti al medio.}{Caparezza\\
Il sogno eretico, 2011} 
\tableofcontents %\listoffigures
\begin{mychapter}[Abstract]
Nanoscopic system in quantum coherent regime are at the center of many 
research fields in physics, from quantum computing and cold atoms to
transport in nanoscopic and mesoscopic systems.

The quantum coherence induces the growth of many interesting features.
In this thesis we focus our attention on two important consequences of
the quantum coherence: Dicke superradiance \cite{dicke} and Anderson localization \cite{anderson}. 

\begin{figure}[h]
\centering
\includegraphics[width=0.6\textwidth]{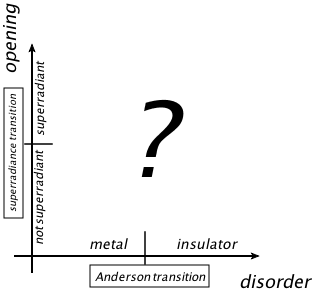}
\caption{Phases diagram of the problem of the interplay of superradiance and disorder.}
\label{fig:phase}
\end{figure} 

Open quantum system can be modeled as a discrete quantum system coupled to an external
environment characterized by a continuum of states.
As a consequence of the coupling to the continuum the eigenstates acquire a decay width (finite lifetime).
For small coupling all the eigenstates acquire a decay width proportional to the coupling.
When a critical value of the coupling is reached the system undergoes a strong change in its resonance structure.
Above this critical value of the coupling some eigenstates 
have non-zero decay width while the decay width of the other eigenstates is approximatively zero.

This phenomena is called transition to superradiance: is due by the opening of the system and induces a segregation of the decay widths of the eigenstates.

On the other hand Anderson transition is driven by intrinsic disorder and induces the exponential localization of the wave functions.

These two phenomena are introduced in Chapter \ref{chap:open} and in Chapter
\ref{chap:anderson} respectively.
The effects described in the seminal papers of R. H. Dicke (1954) and P. W. Anderson (1958) have been extensively studied, separately, in the last fifty year.
On the other hand the interplay between these two phenomena has not been
studied in detail.
In other words we know quite well the effects of a variation of the degree of openness of the system (superradiance transition)
or the degree of disorder (Anderson transition) separately.
What we are going to study is what happens when we vary both the opening and the
disorder.

In Figure \ref{fig:phase} a picture of the problem of the interplay in terms of
phases diagram is shown. 
If we move along the \emph{opening} axis of the phases diagram, at zero disorder we 
reach the superradiant regime and then a segregation of the widths 
occurs, splitting the eigenstates into subradiant and superradiand states.
How these two different subspaces are affected by disorder?

Similarly, if we move along the \emph{disorder} axis at zero opening (closed system), we cross the 
Anderson transition threshold and then all the eigenstates become localized.
How these eigenfunctions are sensitive to the opening?

Of course these questions are very general and, even if we are not able to give a fully exhaustive answer, we have addressed here these issues for a particular
kind of systems.

In particular, in Chapter \ref{chap:1d}, we have studied the $1D$ Anderson model
in presence of coherent dissipation, i.e. where a particle hops from site to site in presence of disorder, and escape to any site is allowed.
This situation occurs when the wavelength of the particle is comparable with the sample size and then an effective long-range hopping is created between the sites.
In this situation, disorder and opening, have opposite effects:
while disorder tends to localize the wave functions, the opening tends to delocalize them, since it induces a long range interaction.
In this system we have characterized the structure of the eigenstates
in different regimes.
The main results is that subradiant and superradiant subspaces
are affected by disorder in a very different way.

In Chapter \ref{chap:3d} we have studied the same problem in a
$3D$ system.
In the $3D$ Anderson model the effect of disorder is very 
different than in $1D$ case since the transition localization-delocalization occurs through a mobility edge.
Nevertheless, the features that we have found in the $1D$ system, turn out to be very general and holds in $3D$ model too.

The understanding of this interplay could also play a crucial role in the quest for Anderson localization of light \cite{qms} and 
matter waves \cite{aspect} even if, due to the generality our theoretical framework, we believe that it can be useful in the description of all those  quantum systems for which
the wavelength of the particle is of the same order of the typical length scale of the system.  

\end{mychapter}

\mainmatter
\chapter{Open quantum systems: superradiance transition}
\label{chap:open}
In this chapter we present the effective Hamiltonian approach to open quantum systems.

The Effective Hamiltonian, $\he$, is a powerful tool to take into account the effect of the coupling of a close discrete
quantum system to a continuum of states which represent the outside world, for example 
the continuum of the modes of the electromagnetic field. 
This method is explained in section \ref{sec:he}.

Transport propieties depend on the degree of openness of the system \cite{celkap, rot, zil, bor} but in important applications
the opening is large and can not be treated perturbatively.
How the eigenvalues of the effective Hamiltonian affect the transport proprieties is shown in
section \ref{sec:smatrix} where the scattering matrix, $S(E)$, in analyzed. 

The analysis of the complex eigenvalues of $\he$ reveals  a general feature of the
open quantum systems: the segregation of the decay widths, i.e. the imaginary part of the eigenvalues, after a critical value of the coupling to the continuum. This phenomena is called
superradiance transition (ST). The name is due by the analogy with Dicke superradiance in quantum optics \cite{dicke}.
In the section \ref{sec:st} this phenomena is explained in detail.

\section{The effective Hamiltonian approach to open quantum systems}
\label{sec:he}
Open quantum system can be modeled as a discrete quantum system coupled to an external
environment characterized by a continuum of states.

In order to approach this problem we split the Hilbert space, $\mathscr{H}$, into two mutually orthogonal subspaces, $S_P$ and $S_Q$.
These subspaces are called \emph{internal subspace} and \emph{external subspace} respectively.
The projection operators of a generic state $\ket\psi \in\mathscr{H}$ on $S_{P/Q}$ subspace are $P$ and $Q$ respectively, so we have
\ba
P\ket\psi &\in& S_P \cr
Q\ket\psi &\in& S_Q.
\ea
The $S_P$ subspace involves the internal states $\{\ket{n}\}$ labeled by a discrete quantum number $n=1, \dots,N$. 
The  $S_Q$ subspace involves the external states $\{\ket{c, E}\}$ labeled by a discrete quantum number $c=1, \dots,M$, 
which represents the decay channel, and a continuum variable, $E$, which represents the energy.

Using the projector operators the full Hamiltonian of the system can be written as
\be
\label{full}
H_f= PHP + QHQ + PHQ + QHP.
\ee
Using Eq.(\ref{full}) and the fact that $\ket\psi = (P+Q)\ket\psi$, the stationary Schr\"odinger equation for the full system read as
\be
\label{full2}
[PHP + QHQ + PHQ + QHP]\ket\psi = E (P+Q)\ket\psi.
\ee
Exploiting the proprieties of the projection operators, namely: $PP=P, QQ=Q, PQ=QP=0$, Eq.(\ref{full2}) became
\be
\label{full3}
(E-PHP)P\ket\psi + (E-QHQ)Q\ket\psi = QHP P\ket\psi + PHQ Q\ket\psi. 
\ee
From multiplication of Eq.(\ref{full3}) by $P$ and $Q$ from the left, we obtain respectively
\ba
\label{full4}
(E-PHP)P\ket\psi&=&PHQ Q\ket\psi \\
&&\cr
\label{full5}
(E-QHQ)Q\ket\psi&=&QHP P\ket\psi.
\ea

Our main purpose is to write the Schr\"odinger equation (\ref{full2}) projected into the internal subspace 
and so we have to eliminate the external states, $Q\ket\psi$.
In order to do this, from Eq.(\ref{full5}) we get 
\be
Q\ket\psi=(E-QHQ)^{-1}QHPP\ket\psi
\ee 
and
we putting it into Eq.(\ref{full4}).
We obtain
\be
\he P\ket\psi = EP\ket\psi,
\ee
where
\be
\he=PHP + PHQ\frac{1}{E-QHQ+\rmi0}QHP,
\ee
is the effective Hamiltonian.
To put it in a more clear form we use the explicit form of the projection operators
\ba
P&=&\sum_n \ket{n}\bra{n},\\
&&\cr
Q&=&\sum_c \int dE' \ \ket{c,E'}\bra{c,E'}
\ea
and multiply from the left by the external state $\bra{m}$.
What we get is
\be
\label{full6}
\braket{m|\he(E)|n}=\braket{m|H|n} + \sum_c \int dE' \ \frac{A_m^c(E')A_n^c(E')^*}{E-E'+\rmi0},
\ee
where $A_n^c(E)\equiv \braket{n|PHQ|c,E}$ represents the coupling amplitudes between the internal and the external states.

The integral in Eq.(\ref{full6}) can be decomposed, using the Sokhotski-Plemelj formula
\footnote{Let $f$ be a complex-valued function, let $a$ and $b$ real 
and $a<0<b$. Then
\be
\lim_{\epsilon\rightarrow0^+}\int dx \ \frac{f(x)}{x\pm\rmi\epsilon} = \mathcal{P}\int dx \ 
\frac{f(x)}{x} \mp \rmi\pi f(0),
\ee
where $\mathcal{P}$ stand for the Cauchy principal value.}, into its Hermitian part (principal value) and the remaining non-Hermitian part,
using
\be
\sum_c \int dE' \ \frac{A_m^c(E')A_n^c(E')^*}{E-E'+\rmi0} = \sum_c \mathcal{P}\int dE' \ \frac{A_m^c(E')A_n^c(E')^*}{E-E'} -\rmi\pi\sum_{c \ (open)} A_m^c(E)A_n^c(E)^*.
\ee

Now we are able to write the effective non-Hermitian Hamiltonian as follow
\be
\he(E)= H_0 +\Delta(E) -\frac\rmi2 W(E),
\ee
where
\ba
(H_0)_{mn}&=&\braket{m|H|n},\\ 
&&\cr
\Delta_{mn}&=&\sum_c \mathcal{P}\int dE' \ \frac{A_m^c(E')A_n^c(E')^*}{E-E'}, \\
&&\cr
W_{mn}&=&2\pi\sum_{c \ (open)} A_m^c(E)A_n^c(E)^*.
\ea

If we restrict our analysis into a small energy windows we can do some useful simplification.
In fact if we assume that $\Delta(E)$ and $W(E)$ are smooth function of $E$ we can neglect their energy dependence.
Therefore the coupling amplitudes, $A_n^c$, becomes energy independent parameters.
For simplicity, in this work, we have neglected the energy shift term, $\Delta$.

In conclusion, the effective Hamiltonian that we use in order to study the coupling of a discrete quantum system to a
environment is:
\be
\label{full7}
\he=H_0 -\frac\rmi2W,\qquad W_{mn}=2\pi\sum_{c=1}^M A_m^c{A_n^c}^*.
\ee
The interpretation of Eq.(\ref{full7}) is the follow:
the Hermitian part, $H_0$, represent the \emph{close} discrete quantum system. The non-Hermitian part, $W$, describes
the coupling to the continuum of $N$ intrinsic states through $M$ open decay channels.

The eigenvalues of $\he$ are complex 
\be
\E_r=E_r -\frac\rmi2\Gamma_r,
\ee
where $\Gamma_r$ is the decay width of the state.
In this approach, the decay width $\Gamma_r$, has to be interpreted as the inverse of a 
characteristic lifetime ($\Gamma_r/\hbar = 1/\tau_r$) of an eigenstate $\ket{r}$.
Indeed it can be proved that the time evolution of $\ket{r}$ is driven by $\he$, so that 
\ba
\label{full18}
\ket{r(t)}&=&\rme^{-\frac\rmi\hbar \he t} \ket{r}\cr
             &=& \rme^{\frac\rmi\hbar E_r t} \ \rme^{-\frac{\Gamma_r}{2 \hbar} t } \ket{r}.
\ea
Equation (\ref{full18}) clearly show that if at $t=0$ the particle is in the state $\ket{r}$ after a time $t$
the probability of find the particle still in this state is 
\be
P_r (t)= \rme^{-\frac{\Gamma_r}{\hbar}t}.
\ee

\section{Transport proprieties: the scattering matrix}
\label{sec:smatrix}
The aim of this section is to show why the transport propriety of the system are strongly affected  by the eigenvalues $\E_r$ of $\he$.
In order to do this we analyze the structure of the scattering matrix, $S(E)$. 
The scattering matrix can be deduced in a straightforward manner from $\he$.
First of all consider Eq.(\ref{full}) and rewrite the full Hamiltonian, $H_f$, as follows
\be
\label{full8}
H_f \equiv H_w + V,
\ee
where $H_w\equiv PHP + QHQ$ is the part of the full Hamiltonian acting within the relative subspace while,  
$V\equiv  PHQ + QHP$, acts across internal and external subspaces.
From standard scattering theory we know that the scattering matrix is defined as
\be
S(E)=1-\te(E),
\ee
where $\te$ is the transmission matrix.
From Eq.(\ref{full8}) and using the Lippmann-Schwinger equation we can write the transmission matrix as
\be
\label{full9}
\te(E)=V + V\frac1{E-H_f+\rmi0}V.
\ee
In accord with Eq.(\ref{full9}) the transition amplitude for the process $b\rightarrow a$ is given by
\ba
\te^{ab}(E) &=& \braket{a,E'|\te|b,E'}\\
&&\cr
&=&\braket{a,E'|V|b,E'}+ \braket{a,E'|V\frac{1}{E-H_f+\rmi0}V|b,E'}.
\ea
Now, using the orthogonality between the two subspaces and the propieties of the projection operators we find
\be
\label{full10}
\te^{ab}(E) =\braket{a,E'| QHP P\frac1{E-H_f +\rmi0}P PHQ |b,E'}.
\ee
The operator $P(E-H_f+\rmi0)^{-1}P$ is the projection in the internal subspace of the full Hamiltonian propagator.
This projection of the full Hamiltonain in the internal subspace is exactly what we have done in the Sec.\ref{sec:he} and leads to effective Hamiltonian.
Of course the projection $H_f\rightarrow\he$ readily leads to the projection of the propagators,
\be
H_f\rightarrow\he \Longrightarrow \frac1{E-H_f}\rightarrow\frac1{E-\he}.
\ee
After this operation the transmission matrix can be written as
\ba
\label{full11}
\te^{ab}(E)&=&\braket{a,E'| QHP\frac1{E-\he}PHQ |b,E'}\cr
&&\cr
&=& \sum_{n,m=1}^N\braket{a,E'| QHP\ket{n}\bra{n}\frac1{E-\he}\ket{m}\bra{m}PHQ |b,E'}\cr
&&\cr
&=& \sum_{n,m=1}^N A_n^{a*}\left(\frac1{E-\he(E)}\right)_{nm}A_m^{b}.
\ea
From the matrix of the transmission amplitude, $\te^{ab}$, we can define the \emph{transmission} from channel $b$ to channel $a$:
\be
\mathsf{T}^{ab}=|\te^{ab}|^2.
\ee

In order to show that the eigenvalues of $\he$ coincide with the poles of scattering matrix, $S=1-\rmi\te$, we have to rewrite Eq.(\ref{full11}) on
the basis of the eigenstates of $\he$.
We start with its diagonalization. Eigenstates of $\he$ form a bi-orthogonal complete set such as
\be
\he\ket{r}=\E_r \ket{r},\qquad \bra{\tilde{r}}\E_r^*=\bra{\tilde{r}}\he,
\ee
with $\bra{\tilde{r}}\neq\ket{r}^*$
and where the eigenvalues are complex 
\be
\E_r=E_r -\frac\rmi2\Gamma_r.
\ee
Hence on the basis of its eigenstates the propagator transforms as
\be
\label{full12}
(E-\he)^{-1}_{nm} \rightarrow (E-\E_r)^{-1}\delta_{r,r'}.
\ee
The coupling amplitudes transform similary
\be
\label{full13}
A_m^b\rightarrow\mathcal{A}_r^b=\sum_{m=1}^N A_m^b \braket{m|r},\qquad A_n^{a*}\rightarrow\tilde{\mathcal{A}}_r^b=\sum_{n=1}^N 
A_n^{a*}\braket{\tilde{r}|n}.
\ee
Armed with Eq.(\ref{full12}) and Eq.(\ref{full13}) the expression for the $\te$ matrix becomes
\be
\label{full14}
\te^{ab}(E) = \sum_{r=1}^N\frac{\tilde{\mathcal{A}}_r^a   \mathcal{A}_r^b}{E - \E_r}.
\ee
It's easy to see, from Eq.(\ref{full14}), that the poles of the $S$ matrix coincide with the eigenvalues of $\he$.
This observation show us that the position in the complex plane of $\E_r$ strongly affects the transport propieties of the system.

\subsection{How to calculate the transmission matrix}
In order to calculate the conductance of the system we need to know the transmission matrix $\te$.
In the previous chapter we have shown that $\te$ is a $M\times M$ matrix, where $M$ is the number of open channels, of the form,
\be
\label{trans}
\te^{ab}(E)=\sum_{m,n=1}^N A_n^{a*}\left(\frac{1}{E-\he}\right)_{nm}A_m^{b},
\ee
The method to calculate the $\te$ matrix naturally implies the diagonalisation of the effective Hamiltonian for any disordered realization and for each energy too if we consider the energy 
dependent formalism.
In this section we will show a new method, developed by S. Sorathia \cite{sorathia}, that involved only the diagonalisation of the Hermitian part, $H$, 
of effective Hamiltonian for any configuration. This is important because the numerical diagonalisation of an Hermitian matrix is much faster than for a non-Hermitian one. 

The effective Hamiltonian can be written as,
\be
\he=H-\frac\rmi 2W.
\ee
where the anti-Hermitian part has the factorised form $W=2\pi AA^T$.
The matrix $A$ is a rectangular $N\times M$ matrix with columns composed of $A_n^c(E)$.
Now we introduce the resolvents, for the closed and for the open system, respectively,
\be
G=\frac1{E-H}, \ \ \ \ \ \ \mathcal{G}=\frac1{E-\he}.
\ee
Since $\he=H-\rmi\pi A^TA$ the relation between $G$ and $\mathcal{G}$ becomes 
\be
\label{res}
\mathcal{G}=G-\rmi\pi G A  \ \frac{1}{1+\rmi \pi K} \  A^T G,
\ee
where we have defined the $M\times M$  matrix $K=A^T G A$. Eq. (\ref{res}) can be easily deduced using the Woodbury matrix identity,
\be
(B+UCV)^{-1}=B^{-1} - B^{-1}U(C^{-1}+VB^{-1}U)^{-1}VB^{-1}.
\ee 
Now we can substitute the relation (\ref{res}) in the definition of transmission matrix (\ref{trans}),
\ba
\label{trans2}
\te &=& A^T \mathcal{G} A \cr
&&\cr
&=&A^T G A -\rmi \pi A^T G A \ \frac1{1+\rmi\pi K} \ A^T G A \cr
&&\cr
&=&K-\rmi \pi K \ \frac1{1+\rmi\pi K} \ K\cr
&&\cr
&=&\frac{K}{1+\rmi \pi K} \  .
\ea
In order to evaluate $K=A^T G A$, and then $\te$, one have to diagonalise the Hermitian Hamiltonian $H$ and to write the matrix $A$  in the basis of $H$.
Using the transformation matrix $V$, which has as columns the eigenstates of $H$, we can write $A$ in the new basis
\be
\tilde{A}=V^T A \Longrightarrow \tilde{A}_n^c=\sum_m A_m^c \phi_m(E_n),
\ee  
where $\phi_m(E_n)$ stands for the $m^{th}$ component of the $n^{th}$ eigenvector of $H$ with eigenvalue $E_n$. 
This change of basis allow us to write the $K$-matrix as
\ba
K^{ab}&=&\left(\tilde{A}^T \frac1{E-H} \tilde{A}\right)^{ab}\cr
&&\cr
&=& \sum_{n=1}^N \ \frac{\tilde{A}_n^a\tilde{A}_n^b}{E-E_n}.
\ea
The $\te$ matrix can be obtained in a straightforward way from Eq.(\ref{trans2}).

In conclusion we can state that the method explained above allow the explicit calculation of $\te$ matrix by the diagonalisation of one internal Hermitian 
Hamiltonian only. 
This is a powerful tool from the numerical point of view and we are reduced to invert the $M\times M$ matrix $1+\rmi\pi K$ for each value of $E$ only. In our models
$M$ is small and this the inversion is very fast.

\section{Transition to superradiance}
\label{sec:st}
Let us start from a simplified version of Eq.(\ref{full7}),
\be
\he=H_0 -\rmi\frac\gamma2W,
\ee
where 
\be
\label{sr2}
W_{ij}=\sum_{c=1}^M A_i^cA_j^{c*},
\ee
$\gamma$ is the parameter controlling the coupling to continuum of states of external world and the basis $\{\ket{i}\}$ are the eigenstates of $H_0$
with eigenvalues $E_i^0$ (and thus $H_0$ is diagonal in this representation with $(H_0)_{ii}=E_i^0$).
We can treat $W$ as a perturbation if $\gamma/D\ll1$ where $D$ is the mean level spacing between neighboring eigenstates of $H_0$. This condition is always 
true if $\braket{\Gamma}/D\ll1$\footnote{
Where $\braket{\Gamma}$ is the \emph{average width} defined as 
\be
\label{avewidth} 
\braket{\Gamma}=\frac1N\sum_{r=1}^N\Gamma_r.
\ee.
Of course $\braket{\Gamma}$ depends of $\gamma$. For example if $W_{ij}=1 \ \forall i,j$ we have 
$\braket{\Gamma}=\gamma$.}.
Under this condition the eigenvalues of $\he$ at the first order in perturbation theory are
\be
\label{sr1}
\E_i=E_{i}^0-\rmi\frac\gamma2W_{ii}.
\ee 
Eq.({\ref{sr1}}) state that when $\braket{\Gamma}/D\ll1$ \emph{all} the state acquire a decay width proportional to $\gamma$.
In the limit $\gamma\gg1$ we have that $W$ is the leading term and $H_0$ can be viewed as a perturbation.
The structure of $W$ (see Eq(\ref{sr2})) allow one to deduce that the rank of $W$ and thus also of $\he$ is $M$, the number of open channels.
From this simple consideration we can state that $W$ has only $M$ non-zero eigenvalues if $M<N$.
Thus in the limit of large coupling to continuum only $M$ out of $N$ states will have a non-zero decay width. These state are called \emph{short-lived} states 
(\emph{superradiant} states).
The decay width of the others $N-M$ states approach zero in this regime, and thus are decoupled from the continuum of states, for this reason they are 
called \emph{long-lived} states (or \emph{subradiant} states).

Summarizing:
for small coupling to continuum, $\braket{\Gamma}/D\ll1$, all states acquire a decay width proportional to $\gamma$ (as predicted by first order perturbation theory).
When the coupling to the continuum reaches a critical value, only $M$ eigenvalues continue to
increase their width while the others $N-M$ eigenvalues start to decrease their width.
Finally, in the limit of large coupling, $\braket{\Gamma}/D\gg1$,
only $M$ states have a non-zero decay width and the decay width of the others $N-M$ states are approximately  zero.
We could say that the system in order to survive to the opening has to rearrange 
itself. The opening then induce a segregation of the imaginary part of the eigenvalues of $\he$.
This phenomenon is called 
\emph{superradiance transition} (ST)\footnote{
The name is due by the analogy with Dicke superradiance in quantum optics.}.
The transition to superradiance is expected to occur for 
\be
\label{sr3}
\frac{\braket{\Gamma}}{D}\approx1.
\ee 
The criterium (\ref{sr3}) for the transition to superradiance is valid in the case of uniform energy and negligible energy shift.

\chapter{Disordered system: the Anderson localization}
\label{chap:anderson}

Ordered structures have been studied since the beginning of Quantum Mechanics.
This is because the periodicity of such materials, and then of the interactions, allow one to employ the Bloch theorem (see Ref.\cite{mermin}). In this way the structure of the eigenstates of rather complicated crystalline materials can be calculated.
In the real world perfect crystals are an exception: disorder, in different degrees, is almost always present.
In solid state physic the disorder can be caused, for example, by few impurities in an otherwise perfect 
crystalline host (weak disorder).
In the opposite limit, if we think about alloys or glassy structure, we have an example of strongly disordered materials.
  
In this chapter we briefly explain the concept of intrinsic disorder in a quantum system and then of the Anderson localization.
The microscopic theory of Anderson localization \cite{anderson} is far from trivial and after fifty years is not fully understood.
However for our purpose is sufficient to understand the consequence of the introduction of the disorder
on the structure of the wave functions.

\section{$1D$ Anderson model}
\label{sec:example1d}
We start from a paradigmatic example that show how the disorder can change drastically the structure of 
the eigenstates of a system.
Consider a one-dimensional Anderson model \cite{anderson, lee} with diagonal disorder in the tight binding approximation, for the motion of a particle in a 
disordered potential.
The Anderson Hamiltonian can be written as
\be
\label{and1}
H_0 = \sum_{j=1}^N E_j \ket{j}\bra{j} \ + \ \Omega \sum_{j=1}^{N-1}(\ket{j}\bra{j+1} + \ket{j+1}\bra{j}),
\ee
where $E_j$ are random variables uniformly distributed in $[-W/2 ; W/2]$
and $\Omega$ is the tunnel transition amplitude to nearest neighbors sites.
This mean that each site , $\ket{j}$, has a random energy $E_j$ but the tunnel transition amplitude to nearest neighbor sites, $\Omega$, remains constant.
Of course, the disorder strength is $W$, and if $W=0$ the system is ordered.
In the case of a ordered chain, $W=0$, the eigenstates are extended
waves
\be
\label{and2}
\psi_q(j) = \sqrt{\frac{2}{N+1}} \ \sin\left( \frac{\pi q}{N+1}j\right),
\ee
and the relative eigenvalues are
\be
E_q=-2\Omega \cos\left(\frac{\pi q }{N+1}\right),
\ee
where $q=1,\dots,N$ is a quantum number and $j=1,\dots,N$ is a discrete coordinate.

If $W\neq 0$ the structure of the eigenstates change drastically:
a $1D$ infinite tight-binding chain with a random diagonal potential will cause \emph{all} eigenstates of the system
to get localized exponentially, even for weak amplitude of fluctuations.
This means that the envelope of the eigenstate $\psi_q(j)$ is centered on some site $j_0$ with the tails 
that approximatively decay exponentially:
\be
|\psi_q(j)|\sim\exp\left( -\frac{|j-j_0|}\xi \right),
\ee
where $\xi$ is the \emph{localization length} of the eigenstate and is a measure of a typical spatial 
extension of the eigenstate.

This fact can be intuitively understood using the perturbation theory at first order:
a zeroth-order description of the eigenstate would be a
bound state or a localized orbital bound by  deep fluctuation in the random potential.
We could then consider the admixture between different orbitals, $\Omega$, as a perturbation. The main point is that such admixture will not produce an extended state composed of linear combinations of infinitely many localized orbitals, as in the case $W=0$. The reason is that orbitals that are close in space, so that the wave functions overlap significantly (in the tight binding approximation only the nearest neighbors), are in general very different in energy, so that the admixture is small because 
of the large energy denominator
\footnote{
At the first order in perturbation theory an eigenstate of the full Hamiltonian (unperturbed Hamiltonian plus a perturbation) is given by the eigenstate of the unperturbed Hamiltonian plus a superposition of all other unperturbed eigenstates that overlap significantly with it ,
weighted over their energy difference.  
}. On the other hand, states that are nearly degenerate are in general very far apart in space, so that the overlap is very small \cite{lee}.

However if the localization length is greater than the system size ($\xi > N$) we say that
the state is extended.
For this reason also in one-dimensional case we can define a size-dependent critical strength of disorder, $W_c$, such 
that $\xi=N$ in a certain position of the band.

\begin{figure}[h]
\centering
\includegraphics[width=0.5\textwidth]{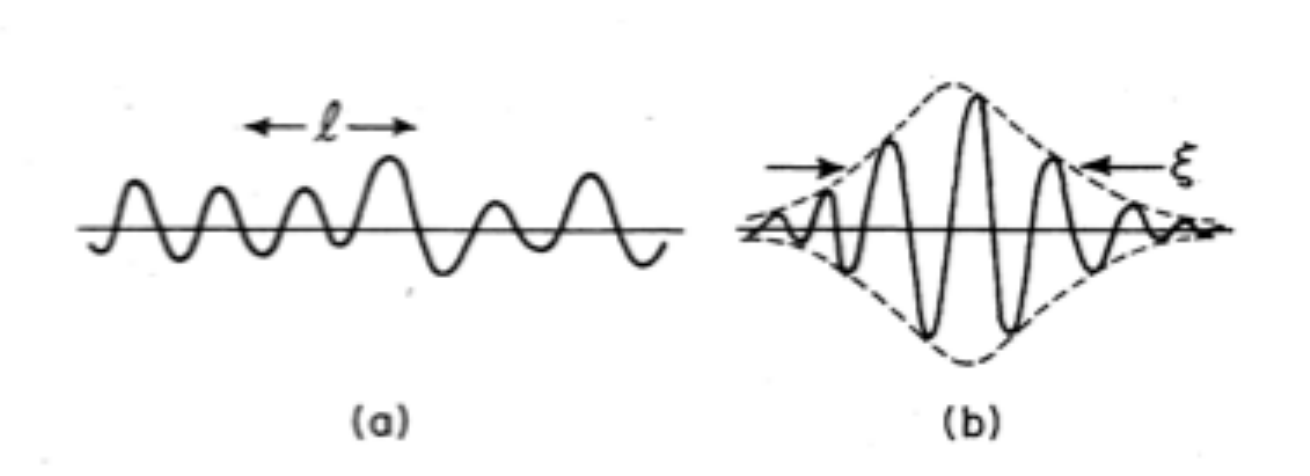}
\caption{(a) Typical extended wave function with mean free path $l$;
(b) localized state with localization length $\xi$.}
\label{fig:loclength}
\end{figure} 

In the case of $1D$ Anderson model with weak uncorrelated diagonal disorder it is possible to find a closed expression for $\xi$. 
Thouless was the first to do it in the case of an infinite chain. He found an expression for $\xi$ in terms of the variance, $\sigma^2$, 
of the site energies distribution \cite{thouless}:
\be
\xi = \frac{8(1-(E/2\Omega)^2)}{\sigma^2} \ \Omega^2.
\ee
In the case of uniform distribution centered around zero we have
\be
\sigma^2=\braket{E^2_j}=\frac{W^2}{12},
\ee
and then the localization length reads as
\be
\label{and3}
\xi=96(1-(E/2\Omega)^2)\left(\frac{\Omega}{W}\right)^2.
\ee
For $E=0$, Eq.(\ref{and3}) has to be modified and we have \cite{felix}
\be
\xi(E=0) = 105.2\dots \ \left(\frac{\Omega}{W}\right)^2.
\ee
Our numerical simulations are in agreement with Eq.(\ref{and3}) in the limit of weak disorder ($\sigma^2\ll1$) 
and far from the band edges ($|E| \ll 2\Omega$).

\section{An overview on $2D$ and $3D$ Anderson model}
\label{sec:anderson}
The number of spatial dimensions strongly affects the phenomenon of localization.
In this section we give an overview of the main results 
about the $2D$ and $3D$ Anderson model.
In both cases the Hamiltonian is similar to (\ref{and1}), i.e. a tight binding model with diagonal uncorrelated disorder. Of course
the interaction $\Omega$ should be extended to nearest neighbors that are $4$ ($6$) in $2D$ ($3D$) case.

\subsection{$2D$ Anderson model}
In a two-dimensional Anderson model we have that for any $W>0$ 
\emph{all} the eigenstates are exponentially localized like in $1D$ case.

\begin{figure}[h]
\centering
\includegraphics[width=0.5\textwidth]{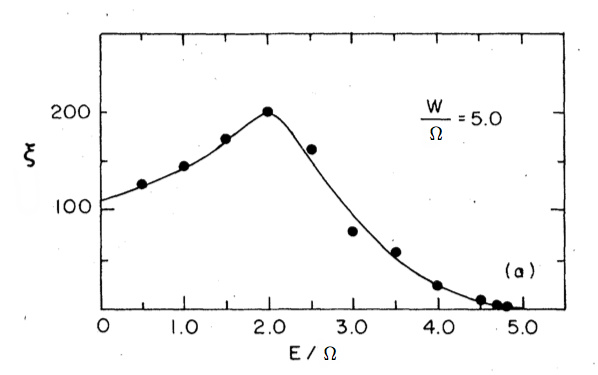}
\caption{Localization length $\xi$ versus energy $E/\Omega$ for a
square lattice.
Full circles are the numerical results while
the thick solid line is the theoretical prediction (taken from Ref. \cite{zdetsis}).
}
\label{fig:zd2}
\end{figure}

\begin{figure}[h]
\centering
\includegraphics[width=0.5\textwidth]{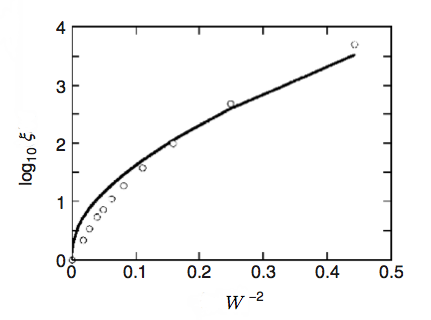}
\caption{Localization length $\xi$ as a function of $1/W^2$ for
the $2D$ Anderson model.
Here is $E=0$.
Open circles are the numerical results while
the solid line is the theoretical prediction.}
\label{fig:xivsw}
\end{figure}

In Figure \ref{fig:zd2} is shown the behavior of $\xi$ as a function of position in the energy band for fixed strength of disorder.
There is no an explicit expression for $\xi$ as a function of all
parameters ($W$ and $E$).
Nevertheless in the center of the energy band and for weak disorder, the localization length behaves as
\be
\label{xiscal}
\ln\xi(E=0) \propto \frac1{W^2}.
\ee
For the detail of derivation of Eq.(\ref{xiscal}) see \cite{sheng}.
In Figure\ref{fig:xivsw} (taken from \cite{sheng}), we show $\log\xi$
vs. $W^{-2}$.
As you can see Eq.(\ref{xiscal}) is verified for small $W$ (large $W^{-2}$)
values.

\subsection{$3D$ Anderson model}
\begin{figure}[h]
\centering
\includegraphics[width=0.5\textwidth]{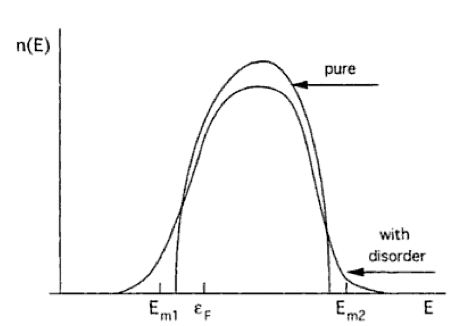}
\caption{Density of states $n(E)$ for $W=0$ (pure) and $W\neq0$ (with dosorder) and the mobility edges in the former case.}
\label{fig:mobility}
\end{figure}

The effect of disorder in the $3D$ model is very different from the $1D$ and $2D$ cases.
For very large disorder all (or almost all) states are localized but for small disorder the situation is different:
the states in the middle of the energy band are extended while the states 
close to the band edges may be localized, see Figure \ref{fig:mobility} \cite{imry}.
The thresholds $E_{m1}$ and $E_{m2}$ are called \emph{mobility edges}.

\begin{figure}[h]
\centering
\includegraphics[width=0.5\textwidth]{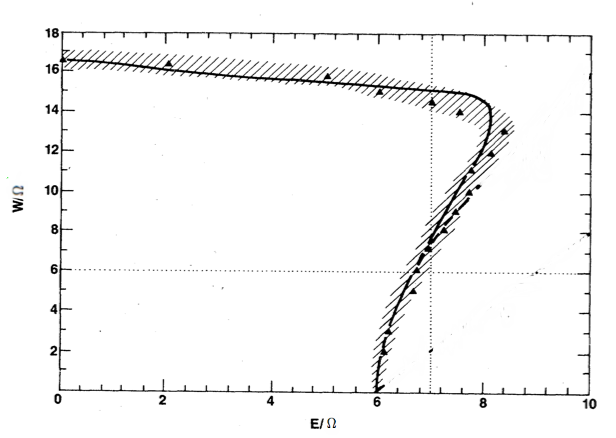}
\caption{Dependence of the mobility edge of diagonal disored $W$ for a cubic lattice. Solid triangles are the numerical results while
the thick solid line is the theoretical prediction.
The dotted straight lines indicate the independent variables for Fig. \ref{fig:xiw}}
\label{fig:zd}
\end{figure}

The position in the energy band of $E_{m1}$ and $E_{m2}$ depends on the ratio $W/\Omega$.
On increasing $W/\Omega$, the mobility edges approach one to each other 
and coalesce at some critical value $W_c$, where all states 
became localized and Anderson transition occurs.
This result was shown numerically for the first time by A. D. Zdetsis \emph{et al.} \cite{zdetsis}, see Figure \ref{fig:zd}.
It is also possible to find that for 
the $3D$ Anderson model 
\be
\label{wcrit}
\frac{W_c}{\Omega} \approx 16.5.
\ee
For finite size sample ($N$) we define the  
\emph{localized regime} when $\xi/N<1$.
 
Close to the critical point, i.e. when $W-W_c\ll1$ and $W>W_c$, the behavior of $\xi$ is given by
\be
\label{xi3}
\xi\propto\frac1{W-Wc}.
\ee
For the details about the theoretical derivation of Eq.(\ref{xi3}) see 
\cite{sheng} and the references therein.

\begin{figure}[h]
\centering
\includegraphics[width=0.9\textwidth]{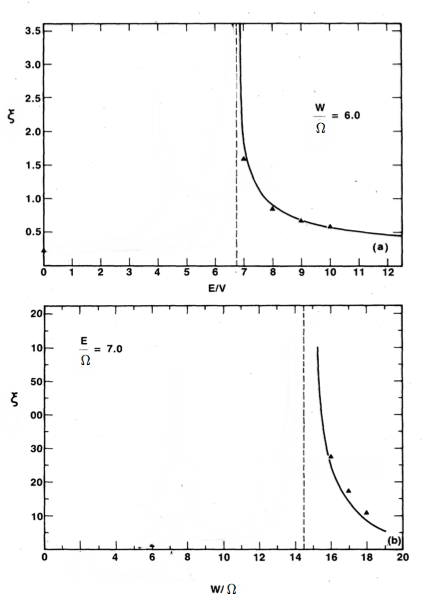}
\caption{(a) Localization length $\xi$ as a function of energy $E/\Omega$ for a cubic lattice with disorder $W/\Omega=6.0$.
(b) Localization length $\xi$ versus the strength of diagonal disorder $W/\Omega$ for a cubic lattice for energy $E/\Omega=7.0$.
In both cases solid triangle are the numerical results while
thick solid line is the theoretical prediction.
}
\label{fig:xiw}
\end{figure}

The localization length, $\xi$, also depends on both the eigenvalues position in the 
energy band and $W$.
Even in this case a closed analytical expression of $\xi$ as a function of both $E$ and $W$ is not know.
However in \cite{zdetsis} the results of numerical simulation
was fitted in order to provide an expression for the localization length:
\be
\label{xi4}
\xi\approx \frac{A\phi + B}{1-\phi} \ l,
\ee
where
\ba
A&=&14.2 \cr
&&\cr
B&=& 2.20\cr
&&\cr
\phi&=&Sl^2/8.96
\ea
where $S$ is the constant energy surface and $l$ is the mean free path
\footnote{
Here both $S$ and $l$ have to be expressed in units of lattice constant
and then are adimensional quantities. 
}.
Using the coherent potential approximation (CPA) to calculate $S$ and $l$
for a given pair of $E$ and $W$ and then substituting into Eq.(\ref{xi4}) an
explicit expression for $\xi$ is obtained.
The localization length, $\xi$, as a function of $W$ and $E$ is shown in Figure \ref{fig:xiw}.

\chapter{Interplay of superradiance and disorder in the $1D$ Anderson model}
\label{chap:1d}

Using a non-Hermitian Hamiltonian
approach to describe open systems, we study the interplay of disorder and 
superradiance in a one-dimensional Anderson model.                         
Analyzing the complex eigenvalues of the non-Hermitian Hamiltonian,
a transition to a superradiant regime is shown to occur.
As an effect of openness the structure of 
 eigenstates  undergoes a strong change in the                       
superradiant regime: we show that the sensitivity to disorder
of the superradiant and the subradiant subspaces is very different;
superradiant states remain delocalized as disorder
increases,  while subradiant states are sensitive to the degree of disorder. 

\section{Introduction}
In this work we analyze a one-dimensional Anderson model, where a particle
hops from site to site in presence of disorder, 
and is also allowed to escape the system from any site.
When the wavelength of the particle is comparable with the sample size,
an effective long-range hopping is created between the sites. 
This coupling can induce the ST, which affects in 
a non-trivial way the transport properties of the system.
Similar models of quantum transport with coherent
 dissipation have been already considered in the literature~\cite{jung1}, but
a detailed analysis of the interplay of
 localization and superradiance is still lacking.
Preliminary investigations have been recently done
 in Ref.~\cite{kaplan, celkap, zil}, but there the particle was allowed to escape only
from the end sites, while in the situation
 analyzed in this work, all sites are coupled to
the external environment.
This situation occurs in many important physical situations, such as
 in cold atoms, where a single photon is injected in the atomic cloud~\cite{kaiser}, or in quantum dots~\cite{exp}. 
 
Intrinsic  disorder and opening to the environment have opposing effects:
while disorder tends to localize
 the wave functions, the opening tends to delocalize
them, since it induces a long range interaction.
The aim of this paper is to study the interplay of disorder and 
opening, and the relation to superradiance.
We show that while below the ST, all states
 are affected by disorder and opening in a similar way,
above it, the effects are quite different for superradiant
 and subradiant subspaces, the latter being more affected
by disorder than the former.

In Sec.~\ref{sec-model} we introduce the model, in Sec.~\ref{sec-st} we analyze the 
ST in our system, and
in Sec.~\ref{sec-results} we present our main numerical results, which we partly
 justify in Sec.~\ref{sec-pt} using perturbation theory.
Finally in Sec.~\ref{sec-conc} we present our conclusions.

\section{Model}
\label{sec-model}

\begin{figure}
\centering
\includegraphics[scale=0.3]{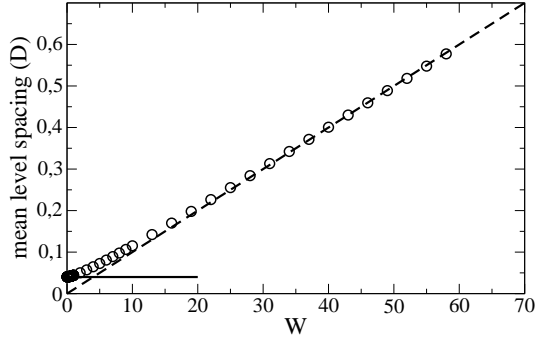}
\caption{The mean level spacing $D$ versus $W$.
The solid line is $4\Omega/N$ while the dashed line is $W/N$.
Each point is obtained averaging over $10^3$ realizations.
Here $\Omega=1$, $N=100$ and $\gamma=0$.}
\label{fig:D_Vs_W}
\end{figure}

Our starting point is the standard one-dimensional Anderson 
model \cite{anderson, lee}, for the motion of a particle
in a disordered potential.
The Hamiltonian of the Anderson model can be written as: 
\begin{equation}
H_0= \sum_{j=1}^{N} E_j | j\rangle \langle j| + \Omega \sum_{j=1}^{N-1} \left(| j \rangle \langle j+1|
+| j+1 \rangle \langle j|\right) \,,
\label{AM}
\end{equation}
where $E_j$ are random variables uniformly distributed
in $[-W/2 ,+W/2]$,  $W$ is a disorder parameter,
 and $\Omega$ is the tunneling transition amplitude
(in our numerical simulations we set $\Omega=1$).
As we have already shown in chapter \ref{chap:anderson}
for $W=0$ the eigenstates are extended and we have for the eigenvalues: 
\be
E_{q} = -2\Omega \cos\left( \frac{\pi q}{N+1} \right), 
\ee
and the eigenstates:
\be
\label{bloch}
\psi_q(j)= \sqrt{\frac{2}{N+1}} \sin \left( \frac{\pi q}{N+1}j \right) \,,
\ee
where $q=1, ...,N$ is a quantum number and $j=1, ...,N$ is a 
discrete coordinate.
In this case, the eigenvalues
lie in the interval $[-2\Omega, 2\Omega]$,
so the mean level spacing can be estimated as $D=4\Omega/N$.
The mean level spacing $D$ as a function of $W$ for the closed model
is shown in Figure \ref{fig:D_Vs_W}.
For $W \ne0$, the eigenstates of the one-dimensional Anderson model
are exponentially localized on the system sites, with exponential
tails given by $|\psi(j)| \sim \exp(-|j-j_0|/\xi)$, and for weak disorder,
the localization length $\xi$ can be written as:
\begin{equation}
\xi \approx 96 \ (1-(E/2\Omega)^2)\left(\frac{\Omega}{W }\right)^{2} \,.
\label{loc}
\end{equation}
For $E=0$, Eq.~(\ref{loc}) has to be modified 
and we have: 
$$\xi \approx 105.2 \ \left(\frac{\Omega}{W}\right)^{2} \,.$$

The phenomenon of Anderson localization
was studied in a closed disordered chain, while
in our case we can vary the degree of openness of the system.
In particular we consider the model 
in which all sites are coupled to a common channel
in the continuum, with equal coupling strength $\gamma$.
This situation can arise when the wavelength of the decaying particle
is much larger than the size of the system. 
This results in a coherent dissipation, which
differs from the usual dissipation where every site decays independently
to a different channel in the continuum. A comparison
between these two different mechanisms will be the subject of a 
future work. 
As we have explained in Chapter \ref{chap:open} the continuum coupling can be taken into account with the aid of an 
effective non-Hermitian Hamiltonian,
which in general can be written 
as,  $$\he (E) = H_0 + \Delta (E) -i Q(E)\,,$$ 
where $H_0$ is the Hermitian Hamiltonian of the closed system
decoupled from the environment and 
$\Delta (E)$ and $Q(E)$ are the induced energy shift 
and the dissipation, respectively.
Neglecting the energy dependence and the energy shift
we have 
\begin{equation}
\he_{ij}=(H_0)_{ij} -\frac{i}{2}  \sum_c A_i^c (A_j^c)^* \,,
\label{Heff}
\end{equation} 
where $A_i^c$ are the transition amplitudes
from the discrete states $i$ to the continuum channels $c$.

In the case under study, we have only one decay channel, $c=1$, and all couplings
are equal, so that
$A_i^1= \sqrt{\gamma}$. Thus
 the effective Hamiltonian can be written as:
\begin{equation}
\he= H_0 -i\frac\gamma2 Q \,,
\label{amef}
\end{equation}
where $H_0$ is the
 Anderson Hamiltonian with diagonal disorder, Eq.~(\ref{AM}),
 and $Q_{ij}=1$ $\forall i,j$.

In order to study the interplay of Anderson 
localization and superradiance we
analyze the participation ratio ($PR$) of the eigenstates 
of $\he$, defined as,
 \be
 PR= \left\langle{\frac{1}{ \sum_i |\langle i| \psi \rangle|^4}}\right\rangle \,,
 \label{pr}
 \ee
 where the average is over disorder.
 
  The $PR$ is a measure of the degree of the spatial extension of wave function.
For example if we consider the completely delocalized state $\ket{\psi}=\frac1{\sqrt{N}}(1, \dots, 1)^T$ we have $PR=N$.
In the opposite limit, for a state localized only on a certain site we have $PR=1$.
All the other kinds of states should have $1 \leq PR \leq N$.
For example, the eigenstates of $H_0$ (closed system) for $W=0$ are the
so called Bloch waves, see Eq.(\ref{bloch}).
These eigenstates are extended but not completely delocalized. If we compute the $PR$ in large $N$ limit we obtain $PR=2N/3$ for any eigenstates \cite{felix}.
Is important to note that in principle the $PR$ does not provide us any information about the structure of a state.
Also the phase correlation between the component of a state 
is completely neglected by the $PR$.

%%%%%%%%%%%%%%%%%%%%%%%%%%%%%%%%%%%%%%%%%%%%

\section{Superradiance transition}
\label{sec-st}
\begin{figure}
\centering
\includegraphics[scale=0.4]{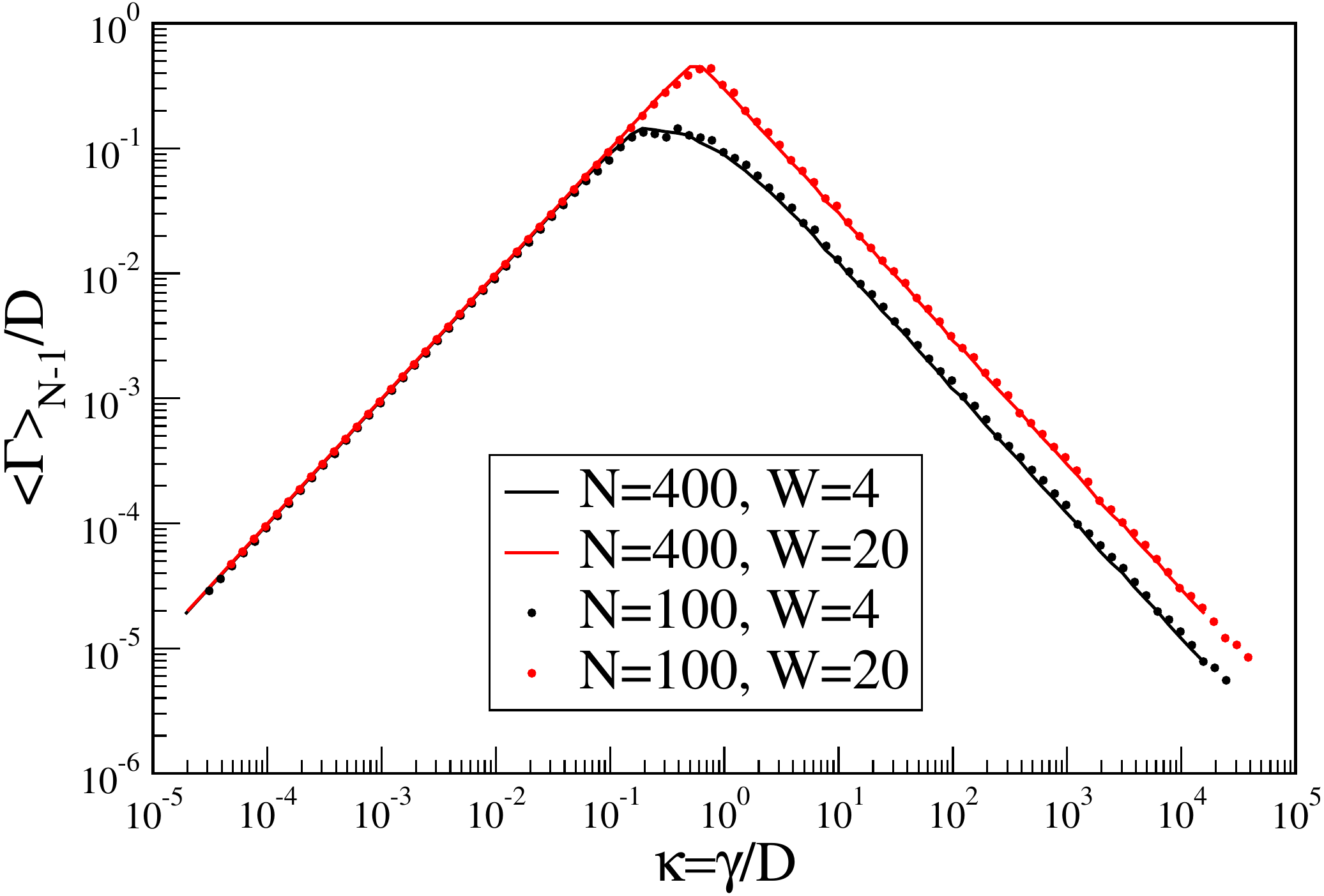}
\caption{The average width of the $N-1$ subradiant states, normalized by the mean level spacing $D$, 
versus the effective coupling 
strength $\kappa$ for different values of $N$ and $W$, and $\Omega=1$.
Here we average over $100$ disordered configurations.}
\label{ave2}
\end{figure}

ST can be analyzed by studying 
the complex eigenvalues 
${\E}_r = E_r -i \Gamma_r /2$ of $\he$
defined in Eq.~(\ref{amef}). 
As the coupling between the  states and the continuum increases, 
one observes a rearrangement of the widths $\Gamma_r$.
ST is expected to occur for $\langle \Gamma \rangle /D \simeq 1$.  
The average width, $\langle \Gamma \rangle$, is given by $\gamma$, so
we can define
\be
\kappa=\gamma/D
\ee
as the effective parameter controlling the 
coupling strength to the continuum. 
In the deep localized regime where disorder is strong ($W \gg \Omega$) we can write $D \approx W/N$, so that the effective coupling strength can be written as:
\be
\kappa=\frac{ \gamma N}{W}
\label{k}
\ee
In Fig.~\ref{ave2} we show that ST
occurs at $\kappa \sim  1$ 
for different values of $W/\Omega$ and $N$.

For $\kappa \gg 1$, we can treat the matrix $Q$ as the leading term in Eq.~(\ref{amef}), and $H_0$ as a perturbation.
The superradiant state $|SR\rangle$ is given to zeroth order by the
eigenstate of $Q$ with nonzero eigenvalue:
 $|d\rangle =\frac1{\sqrt{N}}(1,...,1)^T$, and the energy of $|SR\rangle$
is evaluated at the first order as
\begin{equation}
\langle d| \he|d \rangle =\epsilon -i \frac\gamma2N \,,
\end{equation}
where 
$$\epsilon=\frac1N \sum_{i=1}^N E_i+ 2\Omega \frac{N-1}N$$
 and $E_i$ are the random diagonal elements of $H_0$.
Averaging over disorder and
taking into account that $E_i$ are distributed uniformly
in
  $[-W/2,W/2]$ we obtain,
\begin{equation}
\label{meanen}
\langle \epsilon \rangle=2 \Omega \frac{N-1}N 
\end{equation}
and
\be
\label{variance2}
{\rm Var}(\epsilon) = \langle \epsilon^2\rangle - \langle{\epsilon}\rangle^2 = \frac{W^2}{12N} \,.
\ee
These results agree with our numerical simulations
 for different values of $N$ and 
allow one to know the position in the energy  band
of the superradiant state in the limit $\kappa\gg1$.
From Eq.~({\ref{meanen}}) we deduce that the mean energy 
$\langle{\epsilon}\rangle$ of the superradiant state is independent of $W$.

%%%%%%%%%%%%%%%%%%%%%%%%%%%%%%%%%%%%%%%%%

\section{Numerical Results}
\label{sec-results}
In order to study the interplay of superradiance and disorder we
 have analyzed the $PR$ of the eigenstates
 of the non-Hermitian Hamiltonian, Eq.~(\ref{amef}). 
  
\begin{figure}
\centering
\includegraphics[scale=0.35]{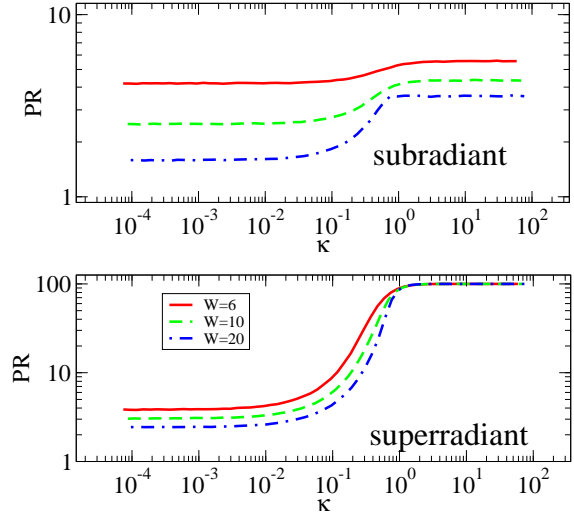}
\caption{The participation ratio $PR$ is shown as a function of $\kappa$ for different disorder strengths. In the upper panel  we consider states with $-1.5 \le E/\Omega \le -0.5$, which
become subradiant for large $\kappa$, while in the lower panel we consider the state with the largest width, which corresponds to the superradiant
state for large $\kappa$. Here $N=100$, $\Omega=1$, and the $PR$ is averaged over $4000$ 
disorder realizations.  }
\label{PRsubsr}
\end{figure}

As explained in the previous section, 
as the coupling with the continuum is increased
we have the formation of one superradiant state (the one with the
largest width) and $N-1$ subradiant ones.
In Fig.~\ref{PRsubsr} (upper panel) we analyze the $PR$ as a function of $\kappa$ for 
the subradiant states for $\kappa > 1$, and in 
 Fig.~\ref{PRsubsr} (lower panel)
we analyze the case of  
the state with the largest width, which becomes superradiant 
for $\kappa > 1$.
As the opening, determined by the parameter 
$\kappa$, increases, the $PR$ of both superradiant and subradiant states
increases, showing that the opening has a delocalizing effect. 
But the consequences of the opening  
are very different for superradiant and subradiant states. 
For the latter, the $PR$ reaches a plateau value above 
the ST ($\kappa \approx 1$), which is 
slightly higher than the $PR$ for $\kappa \ll 1$.  Moreover 
on increasing the  disorder, the $PR$ of the subradiant states
 decreases, both below and above the ST, see Fig.~\ref{PRsubsr} upper panel. 
The situation is different for the superradiant states. 
Above the ST these states become completely delocalized
($PR\approx N$) and the delocalization is not affected by an increase in $W$,
see Fig.~\ref{PRsubsr} lower panel.  

\begin{figure}[h]
\centering
\includegraphics[scale=0.4]{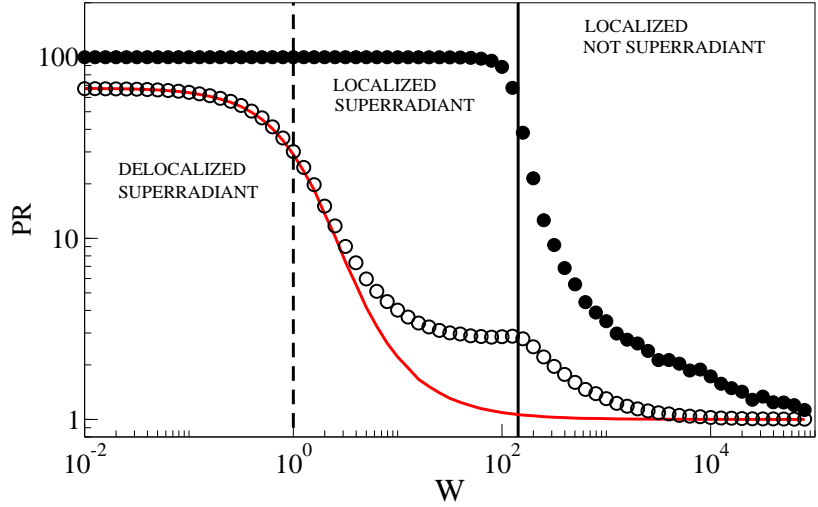}
\caption{The participation ratio is shown as a function of the disorder strength $W$. 
Open  circles stand
 for the subradiant states, full circles indicate
the superradiant state, while the red line stand for the closed system. Each point is 
obtained by averaging over $100$
 disorder realizations for the superradiant state, while for the
subradiant states, an additional average over all the subradiant states
 is performed. 
 For the closed system we have averaged over all the states.
The right and left vertical lines indicate the ST and the
delocalization transition, respectively.
Here $N=100$ and $\gamma=\Omega=1$.
}
\label{PRW3}
\end{figure}

We now look more closely at how the subradiant and superradiant states
are affected differently by increasing the disorder strength $W$. 
In Fig.~\ref{PRW3}, we consider the case of $N=100$ and $\gamma=\Omega=1$. 
For small disorder we have   $D \approx 4 \Omega/N$, so that
 $$\kappa=\gamma/D = \gamma N/4\Omega \approx 25 \gg 1\,.$$
This implies that we are in the superradiant regime.
Moreover, for sufficiently small disorder, we have that the localization length is 
larger than the system size,  $\xi \approx 100 \ \Omega^2/W^2 >N$,
so that both superradiant and subradiant states are delocalized.
 For larger disorder (here  $W>1$) we enter the localized regime, for which
$\xi < N$. In this regime the $PR$ of the subradiant states
decreases, while the $PR$ of the superradiant state remains 
unchanged ($PR=N$),
 signaling a superradiant state that remains completely delocalized. 
 As we increase disorder further, $\kappa$ decreases according to 
Eq.~(\ref{k}). The ST occurs
at $W \approx \gamma N $, here $W \approx 100$, and above this value 
the superradiance effect disappears. 
Summarizing, we have a critical value of disorder ($W \approx 100$
indicated as a full vertical line in Fig.~\ref{PRW3})  separating
the superradiant regime ($\kappa > 1$), from the 
non-superradiant one ($\kappa<1$).   
Only for $W>100$, i.e., below the ST, the superradiant states begin
to localize, and, for very large disorder, corresponding 
to very small $\kappa$,
they behave in the same way as the subradiant states. 

We also note that the subradiant states are affected by disorder as the eigenstates of the closed
system in the delocalized superradiant regime.
When $W>1$ we enter in the localized superradiant regime and the differences between the subradiant modes 
and the eigenstates of the close system become considerable.
This behavior can be viewed as a signature of the fact that the subradiant states are effectively decoupled
from the external world and then they behave similarly to those of the closed system.
For $W\ll1$ the numerical simulation of the $PR$ of the closed system reproduce the analytical prevision $PR=2N/3$.

The regime for which the behavior of superradiant state and
subradiant modes is strongly different is the
\emph{localized superradiant} regime.
The value of $W$ for which Anderson transition takes place
depends on $N$ because
\be
\xi\approx N \Longrightarrow W\approx \frac{10\Omega}{\sqrt{N}}. 
\ee
Let us notice that the value of $W$ for which superradiant transition takes place also 
depends on $N$. According to Eq.(\ref{k}), in order to have $\kappa\simeq1$,
one should put
\be
W\approx\gamma N
\ee
so that $W$ have to increase linearly with $N$.
This means that the range of $W$ where the difference between the 
two subspaces is strong can be increased just increasing $N$:
in Figure \ref{PRW3} the localization transition threshold move to the left and the ST move to the right.

\begin{figure}
\centering
\subfigure[]
{\includegraphics[width=10cm, height=6cm]{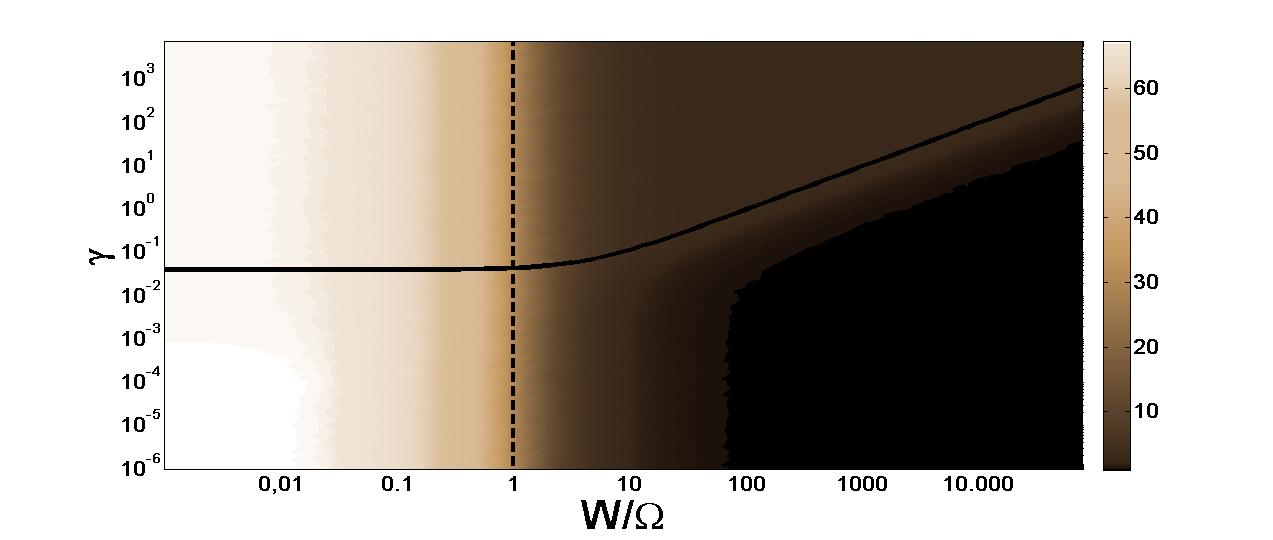}}
\hspace{1mm}
\subfigure[]
{\includegraphics[width=10cm, height=6cm]{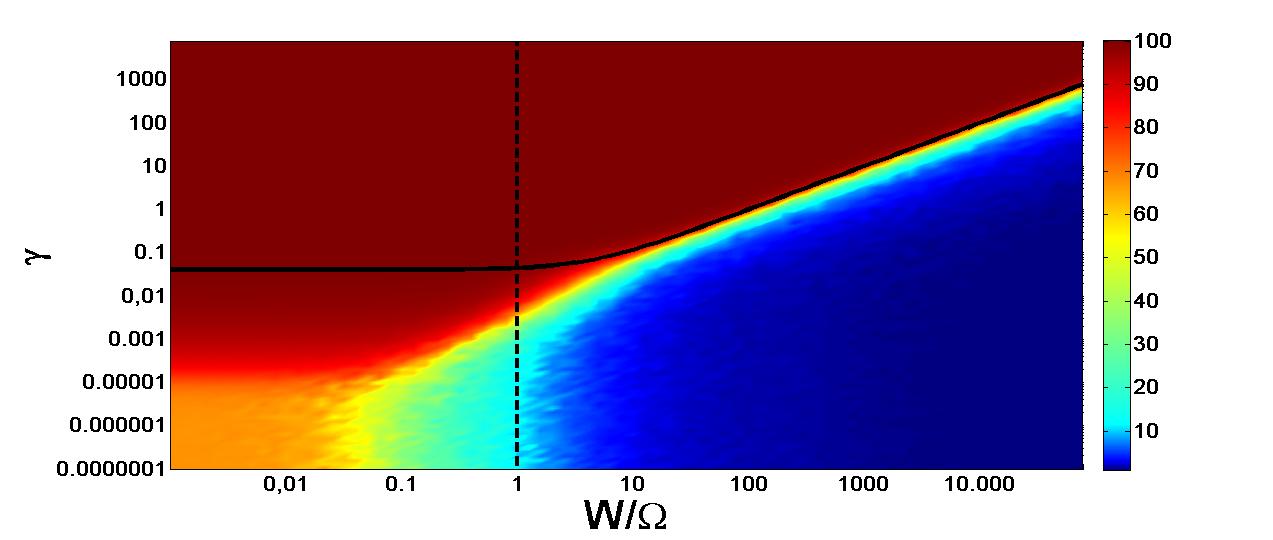}}
\caption{(a) PR of subradiant states as a function of $\gamma$ and $W/\Omega$;
(b) PR of superradiant states as a function of $\gamma$ and $W/\Omega$.
In both cases $N=100$. Each data set is obtained averaging over $10$ realizations for the superradiant state while for the subradiant states an additional average over all the subradiant subspace has been done.}
\label{fig:phases2}
\end{figure}

In this section we have focused our attention on the degree 
of localization of the eigenstates of $\he$ mainly using the PR.
In Figure \ref{fig:phases2} our results are summarized:
the PR is shown as a function
of the opening $\gamma$ and of the disorder strength $W/\Omega$
for a fixed size of the chain.

Note that we use two different colors scale for the two panels in 
order to increase the visibility.

Is important to note that what was known is only the behavior
of the PR for the closed Anderson model, i.e. 
on the $W/\Omega$ axis of the Figure \ref{fig:phases2}.
While here we have fully analyzed the whole $(W/\Omega, \gamma)$ plane.

%%%%%%%%%%%%%%%%%%%%%%%%%%%%%%%%

\section{Discussion}
\label{sec-pt}
In this Section we will justify (using perturbation theory) and briefly discuss the interesting results presented
previously: for small $\kappa$ (below the ST) all the states
are affected in a similar way by the opening and disorder,
 while for large $\kappa$ (above the ST),
the superradiant states remain completely delocalized, independently of
the degree of disorder, while the subradiant states
are still sensitive to disorder, and their $PR$
 decreases with increasing disorder.

\subsection{Perturbative approach for $\kappa \ll 1$}
\label{pertsmallk}

In the limit $\kappa \ll 1$, the eigenstates of $\he$
at first order in perturbation theory can be written as:
\begin{equation}
| n \rangle  =\frac{1}{\sqrt{C_n}}\left[
 | n^0 \rangle  - 
i \frac{\gamma}{2} \sum_{k^0\neq n^0} \frac{\langle k^0 | Q | n^0 \rangle } {E_{n^0} - E_{k^0}} 
|k^0 \rangle  \right] \,,
\label{pert1}
\end{equation}
where $| n^0 \rangle$ are the eigenstates of the closed system, i.e., of the Anderson model.
Of course, the perturbation expansion makes sense only when each coefficient in the
sum in Eq.~(\ref{pert1}) is much less than one. This cannot be true, 
in general, since the eigenvalues $E_{n^0}$ are random numbers uniformly
distributed in the interval $[-W/2, W/2]$.
Thus perturbation theory cannot be applied {\it tout court},
but only for those states whose energies are not too close
one to each other.

\begin{figure}[h]
\centering
\includegraphics[scale=0.4]{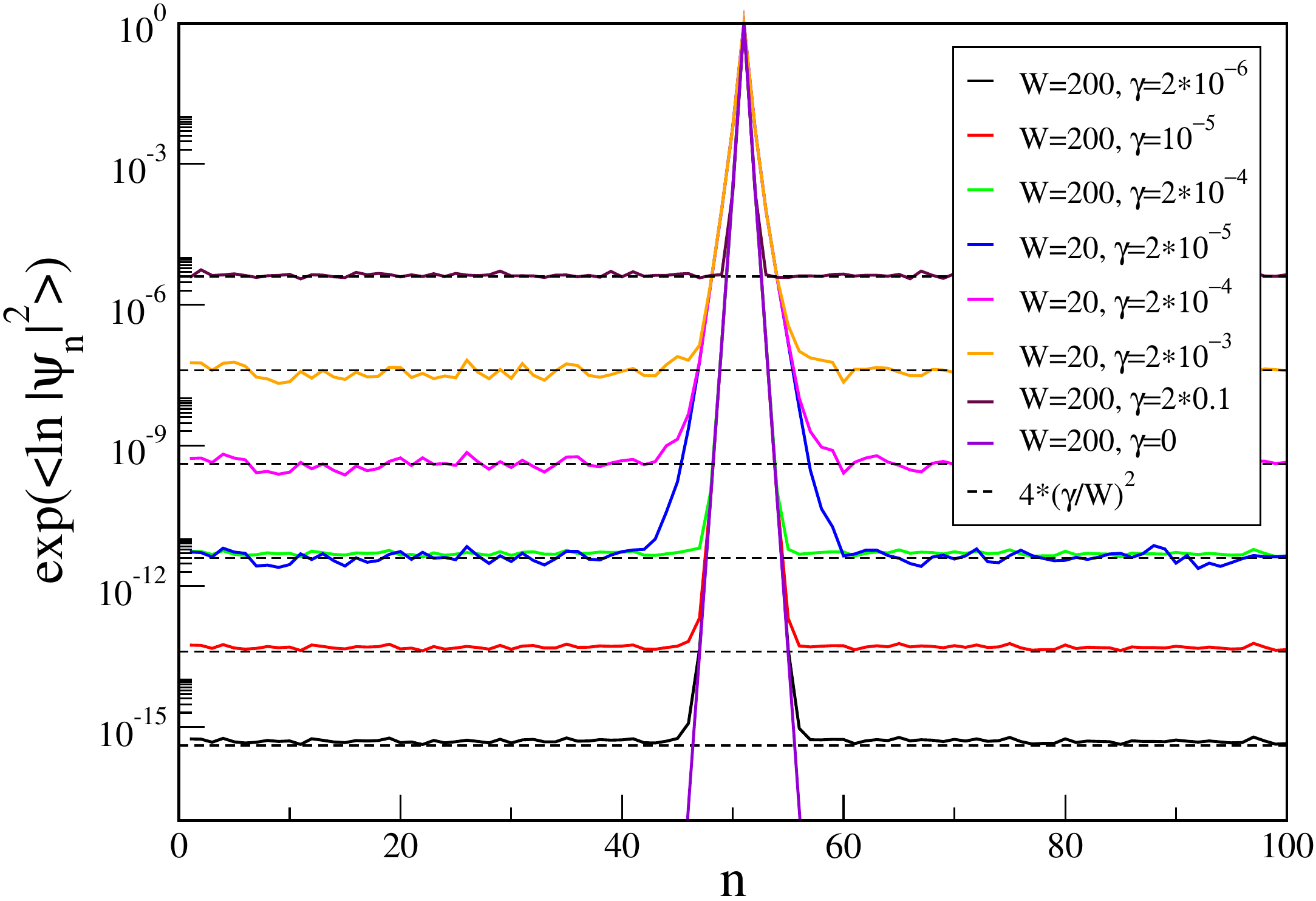}
\caption{The averaged probability distribution of all eigenstates of the non-Hermitian Hamiltonian that are
strongly peaked  in the middle of the chain is shown
for different coupling strength $\gamma$ and disorder strength $W$, as indicated in the caption.
Specifically, we average over all 
eigenstates having a probability $>0.9$ at the site $n=N/2+1$  in order to avoid double-peaked states, and also average over disorder. 
Moreover, to reduce fluctuations, we average the logarithm of
the probability distribution.
In all cases we fix $N=100$ and $\Omega=1$.
Dashed horizontal lines are proportional to $(\gamma/W)^2$ in agreement with
the perturbative approach.
}
\label{PRk}
\end{figure}

This simple observation has deep consequences on the structure
of the eigenstates. Indeed we observed numerically that on the one hand
many single-peaked eigenstates become double- or multiple-peaked
as $\gamma$ increases,
while on the other hand, they all develop a constant plateau
proportional to $(\gamma/W)^2$.
The secondary peak and their positions along the chain are not correlated
and then if we average over the realization we obtain an 
\emph{average probability distribution} which clearly show the plateau
mentioned above, see Fig. \ref{PRk}. 
The physical meaning of the average probability distribution is
discussed in subsection \ref{sec:structure}.

This last fact can be easily explained using first-order perturbation
theory as given by Eq.~(\ref{pert1}): in the deep localized regime 
$W \gg \Omega$, 
the matrix elements $\langle k^0 | Q | n^0 \rangle$ are of order unity and
the average distance between two random energies is $W/3$, 
so that the typical coefficients $\langle k_0|n\rangle$ in Eq.~(\ref{pert1}) 
are $\sim \gamma/W$. 
Furthermore, 
the mean level spacing is $D \approx W/N$, and thus the few largest coefficients 
in Eq.~(\ref{pert1}) are typically $\sim \gamma N/W \sim \kappa$ (using Eq.~(\ref{k})). Thus for weak opening ($\kappa \ll 1$), the {\it typical} eigenstate consists of a single Anderson model eigenstate with a $O(\kappa^2)$ admixture of other states, and therefore the {\it typical} $PR$ for small $\kappa$ differs only by $O(\kappa^2)$ from the $PR$ of the Anderson model.

\begin{figure}
\centering
\includegraphics[scale=0.4]{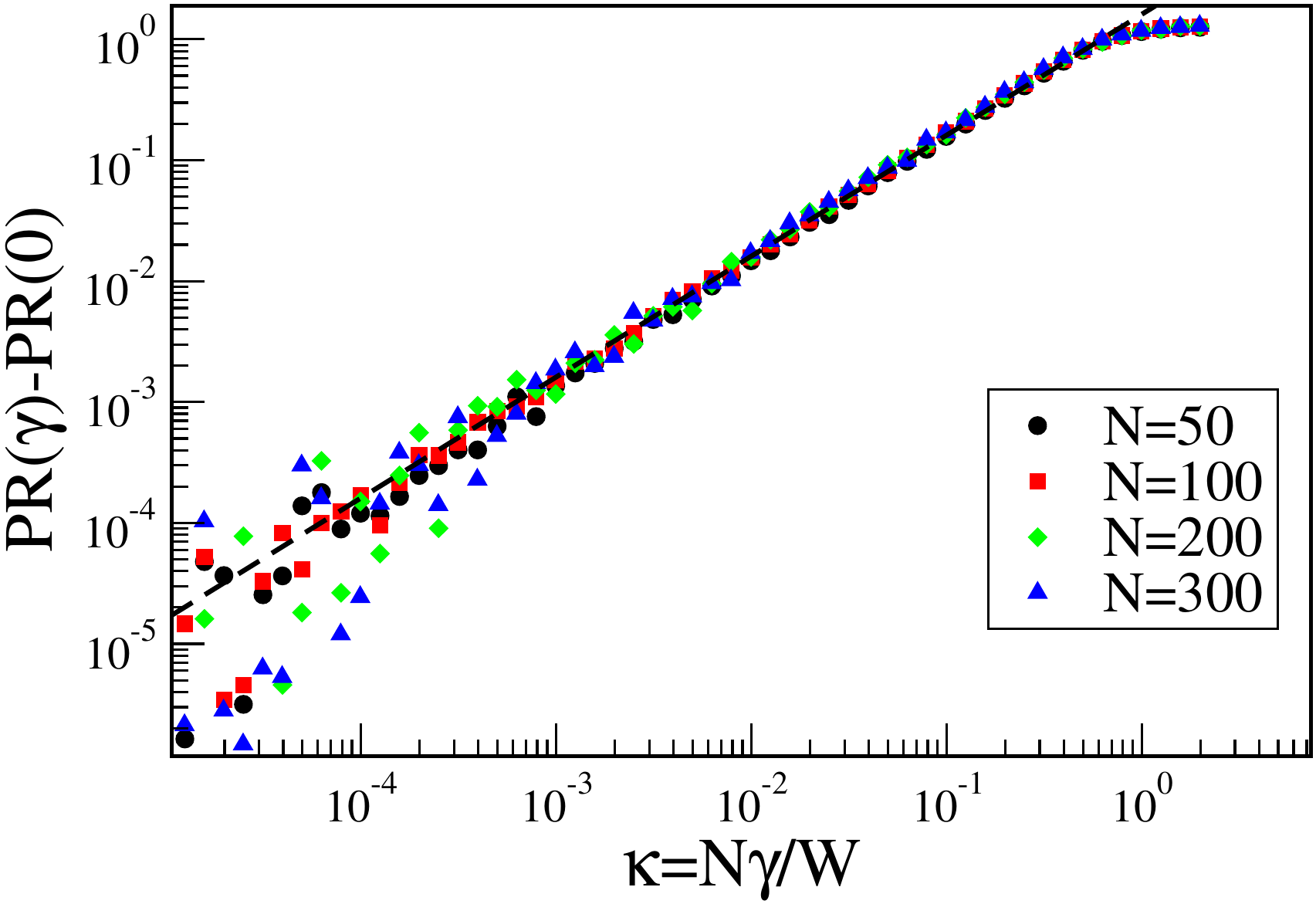}
\caption{The average increase in the participation ratio, compared with the closed
system, is calculated as a function of $\kappa$, for fixed disorder $W=20$ and different system
sizes $N$ as indicated in the legend. In each case the average is performed over 50000 different eigenstates.
The line is $PR(\gamma)-PR(0)=2\kappa$.}
\label{albi}
\end{figure}

As already remarked previously, the perturbative approach
 cannot always work, because for arbitrarily small $\kappa$ 
there is a small but  
finite probability that two energy states are too close 
together. This clustering behavior has important consequences for the localization properties. 
Specifically, since the nearest-neighbor level spacing 
distribution of uniform random numbers $E_{n^0}$ is Poissonian: 
$P(s)=(1/D) \ e^{-s/D}$, where $s$ is the energy difference between 
 nearest-neighbor  levels and $D=W/N$ is the mean level spacing,
we can evaluate the probability to have two levels closer than 
$\gamma/2$ as $1-e^{-\gamma/2D} \approx \kappa/2$ for small $\kappa$. 
This means that there are $\kappa N$ states out of $N$, for which perturbation theory cannot be
applied. When this happens, the Anderson states mix strongly and the $PR$ increases by an $O(1)$ factor. Thus, even though this behavior is rare, it makes an $O(\kappa)$ contribution to the {\it average} $PR$ of the weakly open system, which exceeds the $O(\kappa^2)$ contribution from the typical states.
Indeed the average $PR$ can be evaluated as follow: 
$$
PR= \frac{N\kappa PR_2+ (1-\kappa)N PR_1 }{N}= PR_1 + \kappa (PR_2-PR_1)
$$
where $PR_1$ and $PR_2$ refer to the $PR$ of the states for which
perturbation theory can and cannot be applied.
Since $PR_1 \simeq PR(\gamma=0) + O(\kappa^2)$, and $PR_2 \simeq O(1)$, we have
that  $PR(\gamma) - PR(\gamma=0) \simeq \kappa$.  
The numerical results in Fig.~\ref{albi} confirm that the effect of the opening on the $PR$ grows as $\kappa$, instead of the $\kappa^2$ growth predicted by first-order
perturbation theory.
Here we present the average (over disorder) of $PR(\gamma)-PR(\gamma=0)$,
as a function of 
$\kappa=N\gamma/W$ for fixed disorder strength and different values of the system size.
In any case this is quite a delicate point and we postpone its full analysis
to a future work.

\subsection{Perturbative approach for $\kappa \gg 1$}

In the limit $\kappa \gg 1$ we consider two cases. First, we consider the situation
where the nearest neighbor tunneling 
coupling is $\Omega = 0 $, in which case we can follow the approach explained in Ref.~\cite{sokolov}. 
This approach will be very useful also for the case
$\Omega\neq 0$, which we treat here below. 

\subsubsection{$\Omega=0$ {\rm and} $\kappa \gg 1$}
\label{subsub1}
If $\Omega=0$ the Anderson Hamiltonian is diagonal in the site basis $|j\rangle$ with
 eigenvalues $E_j$ distributed uniformly in the interval $[-W/2,W/2]$.
The eigenstates of the non-Hermitian part $-i\frac\gamma2Q$ of the effective Hamiltonian
are $ | d \rangle  =\frac1{\sqrt{N}}(1,...,1)^T$ (the superradiant state) 
with eigenvalue $-i\frac\gamma2N$, and  $N-1$ degenerate eigenstates $| \mu\rangle$ 
with eigenvalue $0$ (the subradiant states). We will choose $| \mu \rangle $ in a convenient manner later.
Following Ref.~\cite{sokolov} we can rewrite $\he$ in 
the basis of these eigenstates using the transformation matrix $V$, which has
as its columns the eigenstates of $Q$:
\begin{equation}
\tilde{\he} = V^T H_0 V -\rm i\frac\gamma2V^T Q V = \left(
\begin{array}{cc}
%\begin{pmatrix}
 -\rm i\frac\gamma2N  & \vec{h}^T  \\
\vec{h}  & \tilde{H} \\ 
\end{array}
\right ) \,.
\label{h2}
\end{equation}
Here $\vec{h}$ is a vector of dimension $N-1$ with components
\be
h_\mu =\frac1{\sqrt{N}}\sum_{j=1}^N E_j  \langle j|\mu \rangle\,,
\ee
while the matrix elements of the $(N-1)\times(N-1)$ submatrix $\tilde{H}$ are
\be
\tilde{H}_{\mu\nu}=\sum_{j=1}^N E_j \langle \mu|j \rangle \langle j|\nu \rangle.
\ee
Now, we can diagonalize $\tilde{H}$,
\be
\tilde{H}_{\mu\nu}=\sum_{j=1}^N E_j \langle \mu|j \rangle
\langle j|\nu\rangle =\langle \mu|H_0|\nu\rangle = \tilde{\epsilon}_\mu \langle\mu|\nu\rangle.
\ee
Following Ref.~\cite{sokolov} we obtain 
\be
|\mu\rangle =h_\mu \frac1{\tilde{\epsilon}_\mu -H_0} |d\rangle= \frac{h_\mu}{\sqrt{N}}\sum_{j=1}^N \frac1{{\tilde{\epsilon}_\mu-E_j}} |j \rangle\,,
\label{sub}
\ee
where the normalization coefficients $h_\mu$ are given by
\be
h_\mu=\left(\langle d | \frac{1}{(\tilde{\epsilon}_\mu-H_0)^2} |d\rangle \right)^{-1/2} \,.
\ee
In the limit $\kappa \gg 1$, the eigenstates $|\mu\rangle$ of the non-Hermitian part of 
$\he$ are also eigenstates of $\he$.
Since $\langle d|\mu \rangle =0$ we have,
\be
\sum_{j=1}^N\frac{1}{\tilde{\epsilon}_\mu-E_j}=0.
\ee
Therefore  each eigenvalue of $\tilde{H}$ lies between two neighboring
 levels $E_n$, so that the values $\tilde{\epsilon}_{\mu}$ are also confined
in the interval  
$[-W/2, W/2]$. 

Let us now estimate the magnitude of the
mixing matrix elements $h_\mu$. To do this we compute
\be
\vec{h}\cdot\vec{h} = \frac1N\sum_{\mu=1}^{N-1} \sum_{i=1}^N \sum_{j=1}^N \ E_i E_j
\langle \mu|i \rangle \langle j|\mu\rangle \,,
\ee
and using the completeness relation $\sum_{\mu=1}^{N-1} \langle j|\mu \rangle 
\langle \mu|i \rangle 
=\langle j|i \rangle-1/N$ we have
\be
\vec{h}\cdot\vec{h} =  \langle E^2\rangle -{\langle E\rangle}^2 = \Delta{E}^2 \,.
\ee
This leads to
\be
|h_\mu|\sim \frac{\Delta E}{\sqrt{N-1}} = \frac{W}{\sqrt{12(N-1)}}\,.
\ee

Each eigenstate $| \mu \rangle $ in Eq.~(\ref{sub}) is a  superposition of all the site states $ |j \rangle $
with amplitudes $\frac{h_\mu}{\sqrt{N}(\tilde{\epsilon}_\mu-E_j)} \sim \frac{W}{N(\tilde{\epsilon}_\mu-E_j)}$ that depend
only on the energies $E_j$ and not on the site positions $j$.
Nevertheless, each state $|\mu\rangle$ is 
quite localized, since the amplitudes are of order unity for the $O(1)$ number of sites whose energy is
within a few mean level spacings of $\tilde{\epsilon}$ (i.e., when $|\tilde{\epsilon}_\mu-E_j |\sim D =W/N$), and small otherwise.
This small value of the $PR$ for the subradiant states should be compared 
with $PR=N$ of the superradiant states.

The values obtained above for the subradiant and the superradiant states
 correspond to zeroth-order perturbation theory. 
On the other hand
first-order perturbation theory gives:
\begin{eqnarray}
&&| SR \rangle = \frac{1}{\sqrt{C}}\left[ |d \rangle
 + \frac{W}{\sqrt{12(N-1)}} \sum_{\mu=1}^{N-1}\frac{r_\mu}{-
i\frac\gamma2N -\tilde{\epsilon}_\mu} | \mu \rangle\right]\cr  
&&\cr
&&=\frac{1}{\sqrt{C}}\left[ |d \rangle - 
  \ \frac1{\kappa \  \sqrt{3(N-1)}} \sum_{\mu=1}^{N-1}\frac{r_\mu}{i+2\tilde{\epsilon}_\mu/\gamma N} 
| \mu \rangle\right]\cr
&&\cr
&&| {SU\!B}_{\mu}\rangle= \frac{1}{\sqrt{C'_\mu}}\left[| \mu \rangle + \frac{W}{\sqrt{12(N-1)}} 
\frac{r_\mu}{\tilde{\epsilon}_\mu+i\frac\gamma2 N} |d \rangle\right]\cr
&&\cr
&&=\frac{1}{\sqrt{C'_\mu}}\left[| \mu \rangle +
 \frac1{\kappa \sqrt{3(N-1)}}\frac{r_\mu}{i+2\tilde{\epsilon}_\mu/ \gamma N} | d \rangle
\right]\,,
\label{firstorder}
\end{eqnarray}
where $r_\mu$ are random coefficients with $\langle r_\mu^2\rangle=1$. 
We see that  the exact superradiant state $| SR \rangle$ is a combination of the 
unperturbed superradiant state $| d \rangle$
 and a small admixture of the unperturbed subradiant states $| \mu \rangle$, and the mixing 
probability decreases as $1/ \kappa^2$ for large $\kappa$. Similarly, the
admixture of the unperturbed superradiant state $| d \rangle$ in
each exact subradiant states $| {SU\!B}_{\mu}\rangle$ decreases as  $1/ (\kappa^2 N)$.
This shows that $PR \approx N$ for the superradiant state and $PR \sim 1$ for the subradiant
states when $\kappa \gg 1$.

\subsubsection{$\Omega \ne 0$ {\rm and} $\kappa \gg 1$}

As a first step we write the Anderson Hamiltonian $H_0$
in terms  of its eigenstates $|n \rangle$.
Obviously the form of $| n \rangle$ will depend on the degree of disorder $W$.
In the following we limit our considerations to the large disorder regime,
so that in the basis of the eigenstates of  $H_0$, 
the matrix elements of $Q$ remain of order one, $Q_{nm} \sim 1$, 
and we can use the results of Sec.~\ref{subsub1}, with the
site states and energies $|j\rangle$ and $E_j$ replaced by the
Anderson eigenstates and eigenenergies $|n\rangle$ and $E_n$.

In Fig.~\ref{PRW3} we see that for $\kappa>1$ (corresponding to
$W<100$), the superradiant state remains unaffected by the increase 
of disorder, while the subradiant states become more localized 
as the disorder strength is increased.  
The results of the previous section can be used to understand this strongly 
asymmetric behavior of the $PR$ between the subradiant states and the 
superradiant state. 
Indeed at zeroth order in perturbation theory we can see that the
 superradiant state $|SR\rangle \approx |d\rangle$ is completely delocalized, $PR= N$, while
 subradiant states $| {SU\!B}_{\mu}\rangle \approx |\mu\rangle$
 become more and more localized as we increase disorder.
Specifically, the site states $|j\rangle$ in Eq.~(\ref{sub})
are replaced with Anderson eigenstates $|n\rangle$, with localization length  $ \xi \propto 1/W^2$.  
This difference persists in first-order perturbation theory, since the mixing probability between the super- and sub-radiant states
decreases as $1/\kappa^2$ for large $\kappa$, see Eq.~(\ref{firstorder}).

Our perturbative approach justifies the results presented in Fig.~\ref{PRW3},
where we  can see  that 
the subradiant states become increasingly localized 
as we increase disorder.   
At the same time Fig.~\ref{PRW3} shows that
the superradiant state remains completely delocalized on increasing $W$, until
we reach the value $W\approx142.8$ ($\kappa=0.7$) where we find 
numerically that the ST
takes place. 
The perturbative approach shows
 that superradiant states are much less sensitive to disorder
because their complex energies
are at a distance greater than  $\gamma N/2 = W \kappa/2 $  
from the subradiant states. 

%%%%%%%%%%%%%%%%%%%%%%%%%%%%%%%%%%%%%%%
\section{Further results}

\subsection{Structure of the averaged probability distribution in large disorder limit}
\label{sec:structure}
In this subsection we show some preliminary results on the structure
of the \emph{averaged probability distribution} (APD) of the eigenstates for $W/\Omega\gg1$ in the limit of small ($\kappa\ll1$) and large ($\kappa\gg1$) opening. 

We recall that if $\psi$ is a wave function, the averaged probability distribution is defined as $\braket{|\psi|^2}$, where the symbol $\braket{\dots}$ stand for the average over different random
realizations.

Sometimes, in order to reduce fluctuations, we made a 
logarithmic average, i.e. $\exp(\braket{\ln\psi})$, instead of the standard average.

The two different average method, classical and logarithmic, can give generally 
different results.
Is important to note that if $\psi$ is normalized the APD is normalized too only if a classical average is made, otherwise it is not.

The study of the APD is very important because it is a quantity which can be
measured experimentally.
In some interesting recent papers, see for example \cite{aspect, bec}, the diffusion of a
non-interacting Bose-Einstein condensate, in a one-dimensional
disordered potential is measured in order to study Anderson localization.
The quantity measured in this kind of experiments is the spatial distribution
of the atomic density (number of atoms / $\mu m$).
If the particles that constitute the condensate are non-interacting then
the APD can be related to the atomic density profile.

As we have shown in Figure \ref{PRk}, in the limit $\kappa\ll1$ and $W/\Omega\gg1$, the APD
of the eigenstates peaked in a certain site, consists of two terms: the APD 
of the closed Anderson model  and a plateau proportional to $(\gamma/W)^2$ and independent of $N$.
This result can be explicitly deduced using first-order
perturbation theory,
see Eq.(\ref{pert1}).

In the opposite limit, $\kappa\gg1$ and $W/\Omega\gg1$, we have found
a very similar APD for the eigenstates 
of $\he$.
In this regime, the APD of the eigenstates peaked in the middle of the chain consists of two terms: the APD of the closed Anderson model and a plateau
proportional to $1/N$ (and independent of $\gamma$ and $W$).

\begin{figure}
\centering
\includegraphics[width=1.0\textwidth]{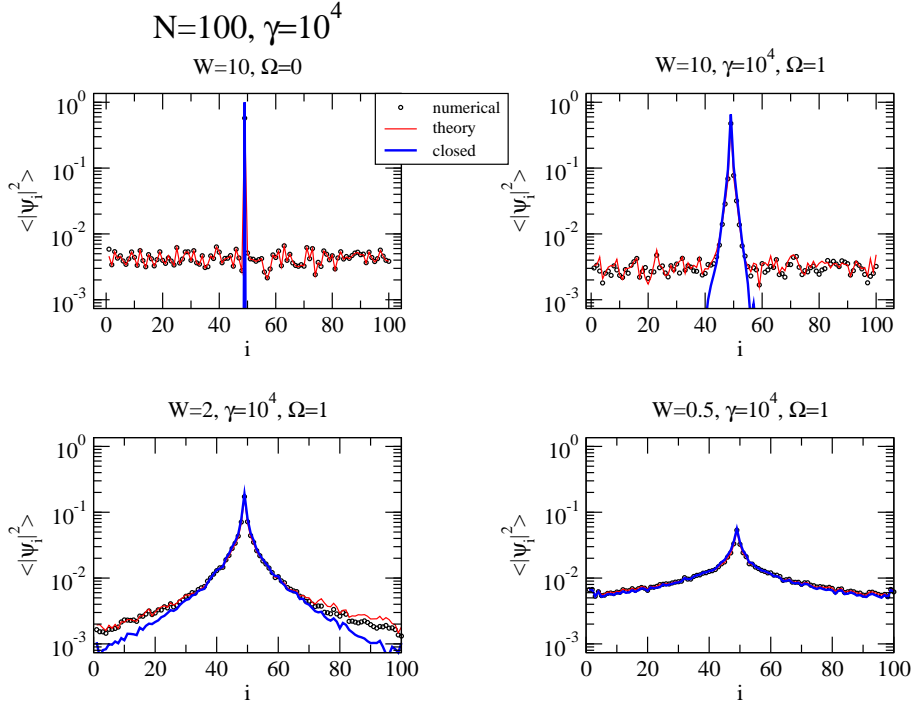}
\caption{Comparison between perturbative formula (\ref{sub}) and 
numerical results for different $W$.
Here an average over $10^3$ disordered configuration is made.}
\label{fig:pertklarge}
\end{figure}

Also in this case first-order perturbation theory can predict the structure of the wave function in $\kappa\gg1$ limit, see Eq.(\ref{sub}).
Even if we have not been able to deduce explicitly the value of the plateau
from Eq.(\ref{sub}) we have checked numerically that Eq.(\ref{sub})
can reproduce the structure of the eigenstates in this limit, 
see Figure \ref{fig:pertklarge}.
It is important to note that Eq.(\ref{sub}) correspond to zero-th order 
expansion of the subradiant eigenstates of $\he$.
In principle Eq.(\ref{sub}) is valid in the large disorder limit, however from
Figure \ref{fig:pertklarge} we can see that it is valid also for 
$W=2$ and $W=0.5$.

\begin{figure}
\centering
\includegraphics[scale=0.6]{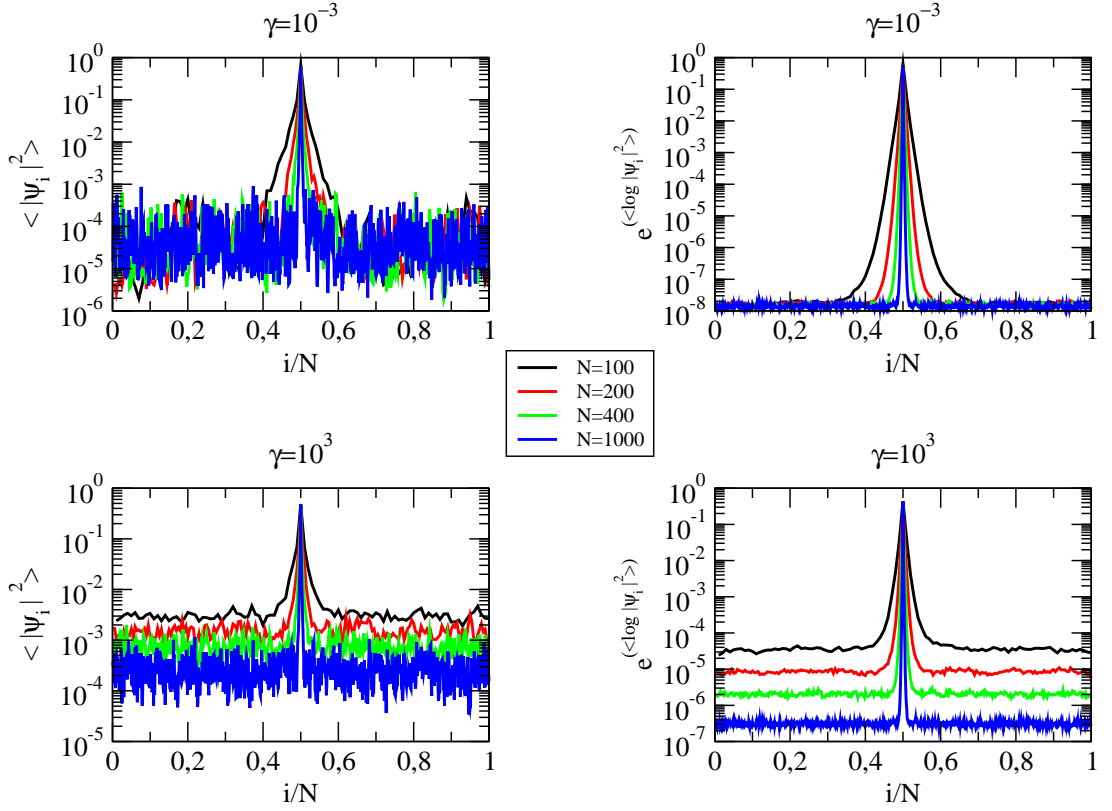}
\caption{In the upper panels we compare the two kinds of average in the limit $\kappa\ll1$ and $W/\Omega\gg1$. In this case the plateau are independent of $N$ and so the averages give the same result: the plateau does not scale with $N$, but we note also that the absolute value of the plateau is different. In the lower panels we compare the two average in the limit $\kappa\gg1$ and $W/\Omega\gg1$. We can see that in the case of classical average the plateau scale as $N$ while in the case of logarithmic average the plateau scale as $N^2$. Here we have set $W=10$ and we average over
$10^3$ configuration.
}
\label{fig:bigkplateauaverage}
\end{figure}

A comparison between these two structures is shown in Figure \ref{fig:bigkplateauaverage}.
In this figure we can also compare the results of the two different kind 
of average, classical and logarithmic.

In the large disorder limit the localization length is $\xi\approx1$.
For this reason, if we compute the APD (from a classical average)
of the eigenstates peaked in the middle of the chain, $\ket{\psi}$,  the normalization condition for the APD can be written as
\be
\label{norma}
A+(N-1)B=1,
\ee  
where $A=\braket{|\psi_{N/2}|^2}$ and $B=\braket{|\psi_i|^2}$ with $i\neq N/2$. 
Of course here $A$ and $B$ are the values of the APD
of the top of Anderson peak and of the plateau respectively.

As a consequence of Eq.(\ref{norma}),
the different behavior of the plateau in the regimes discussed above
implies a different behavior of the Anderson peak.
In fact from Eq.(\ref{norma}) is possible to deduce that
\be
A=1-(N-1)B.
\ee
In the $\kappa\ll1$ limit, we have $B\approx\beta(\gamma/W)^2$
and then for large $N$,
\be
A\approx1-\beta \ N \left(\frac{\gamma}{W}\right)^2.
\ee
In the $\kappa\gg1$ limit, we have $B=\alpha/N$ and then
for large $N$,
\be
A\approx1-\alpha.
\ee
Here $\alpha$ and $\beta$ are two constants.

Summarizing the coefficients $A$ and $B$ have to respect the relation
\ba
\label{ab}
\kappa\ll1\Longrightarrow A&\approx&1-\beta \ N \left(\gamma/W\right)^2 \cr
B&\approx&\beta(\gamma/W)^2 \cr
&&\cr
\kappa\gg1	\Longrightarrow A&\approx&1-\alpha \cr   
B&\approx&\alpha/N.
\ea

\begin{figure}[H]
\centering
\includegraphics[width=1.1\textwidth]{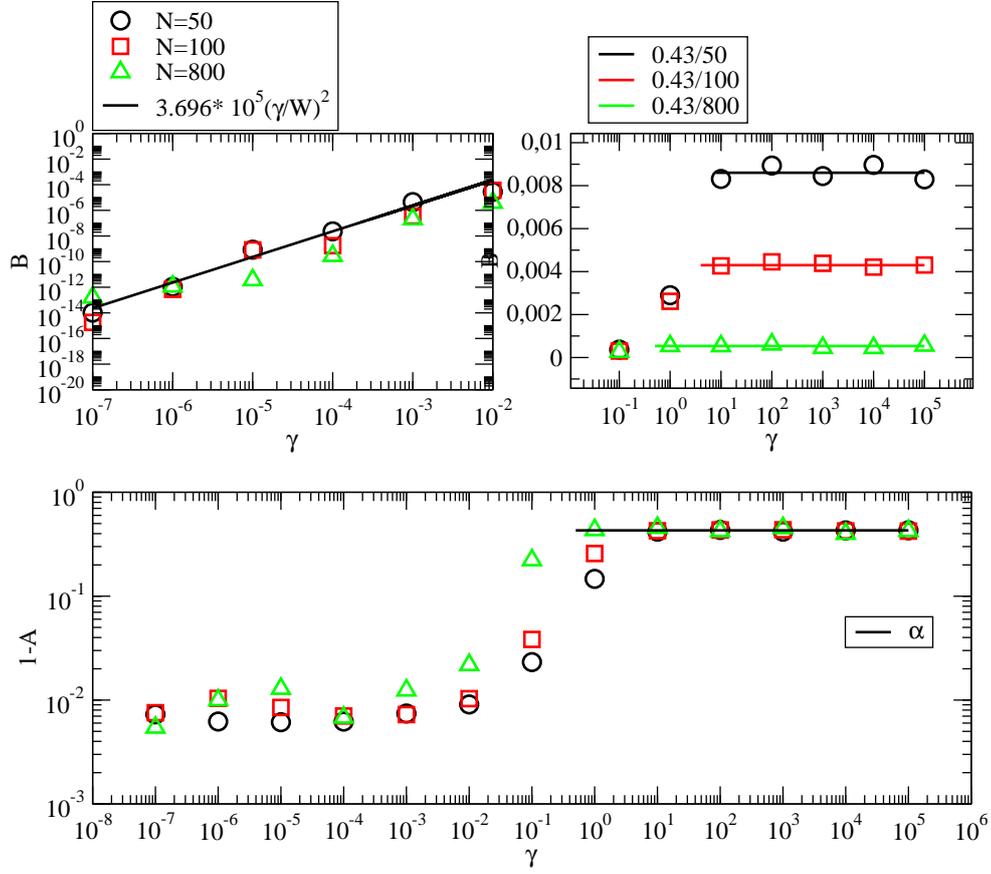}
\caption{Upper panels: $B$ as a function of $\gamma$.
Lower panels: $1-A$ as a function of $\gamma$.
Left figures are for $\kappa\ll1$ while right are for $\kappa\gg1$.
The symbols are the numerical results while the solid lines are Eqs. (\ref{ab}).
The free parameters $\alpha$ and $\beta$ are obtained fitting the
data relative to $B$ and used to compute the theoretical formulae
for $1-A$.
Each point is obtained averaging over $10^3$ realization for $N=50$,
$100$ realization for $N=100$ and $10$ realization for $N=800$.
Here $W/\Omega=400$.
For each value of $N$ the ST takes place for a different value of $\gamma$, see Eq. (\ref{k}).
In the right upper panel, the horizontal solid lines, starts for 
$\kappa=1$. 
}
\label{fig:ab}
\end{figure}

In Figure \ref{fig:ab} we have compared Eqs. (\ref{ab}) with the numerical results.
The coefficients $\alpha$ and $\beta$ are obtained fitting the data relative to $B$, see upper panels.
From the fit of numerical results we obtain
\ba
\alpha&\approx&0.43\cr
&&\cr
\beta&\approx&3.7\cdot10^5.
\ea

The theoretical previsions \ref{ab} are not in agreement with 
the numerical results for the $A$ coefficients in the limit
of small coupling ($\kappa\ll1$).
This discrepancy arise from the rough approximation of the APD, see Eq(\ref{norma}).
In order to avoid this problem a more accurate description of the APD is needed:
we have to take into account the exponential shape of 
the Anderson peak.

\begin{figure}[H]
\centering
\includegraphics[width=1.1\textwidth]{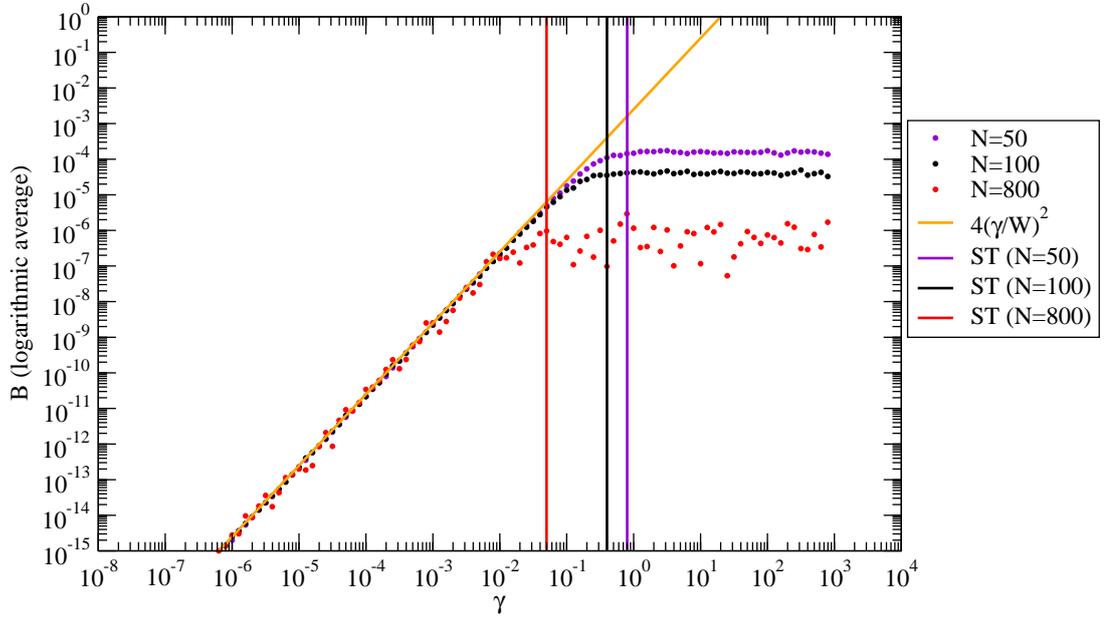}
\caption{$B$ as a function of $\gamma$.
The circles are the numerical results while the solid lines are the theoretical prevision.
Each point is obtained averaging over $10^3$ realization for $N=50$,
$100$ realization for $N=100$ and $10$ realization for $N=800$.
Here $W/\Omega=40$.
The vertical solid line are the STs for different $N$, the colors reflect the different numerical data.
}
\label{fig:plateau1d}
\end{figure}

In Figure \ref{fig:plateau1d} we focus our attention on the plateau, $B$.
Here $B$ is plotted as a function of $\gamma$ for $W=40$ and different
size of the chain.

From this figure we clearly note that the ST is approximatively the threshold where we have a discontinuity in the plateau behaviors.
This is not trivial because we have considered only the cases $\kappa\ll1$ and $\kappa\gg1$ and in principle we do not know the behavior of $A$
and $B$ for intermediate values of $\kappa$.
The results confirm that, when 
a logarithmic average is made, 
the plateau increase as $B\approx4(\gamma/W)^2$ for $\kappa\ll1$
while for $\kappa\gg1$ the scaling law is $B\propto 1/N^2$.

\subsection{Conductance}
In order to understand how the opening affects the transport propieties 
of the system we have studied the conductance.
\begin{figure}[h]
\centering
\includegraphics[width=0.6\textwidth]{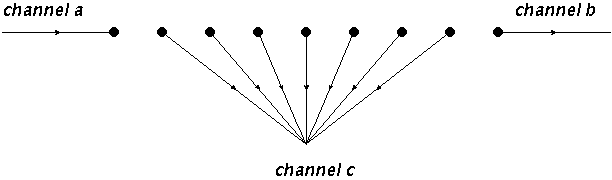}
\caption{Modify model used to compute the conductance.}
\label{fig:cond}
\end{figure}
The modified model used to compute the conductance is shown in Figure \ref{fig:cond}. The edge state on the left, $1$, is coupled to channel $a$ with coupling
strength $\gamma_a$, the edge state on the right, $N$, is coupled to channel $b$ with coupling strength $\gamma_b$, and the sites in the middle, between $2$ and $N-1$, are coupled to a common channel $c$ with coupling strength $\gamma_c$. 
In this framework one should treat the $a$-channel  and the $b$-channel as \emph{left} and \emph{right} channels corresponding to incoming and outgoing waves respectively.
The conductance from $a$ to $b$ channel is given by the adimensional Landauer formula \cite{beenakker, datta} in the
standard way
\ba
G(E)&=& \mathsf{T}^{ab}(E)\cr
&=&|\te^{ab}(E)|^2,
\ea
where $\te$ is the transmission matrix defined in Eq.(\ref{full11}).
\begin{figure}
\centering
\includegraphics[width=1.4\textwidth]{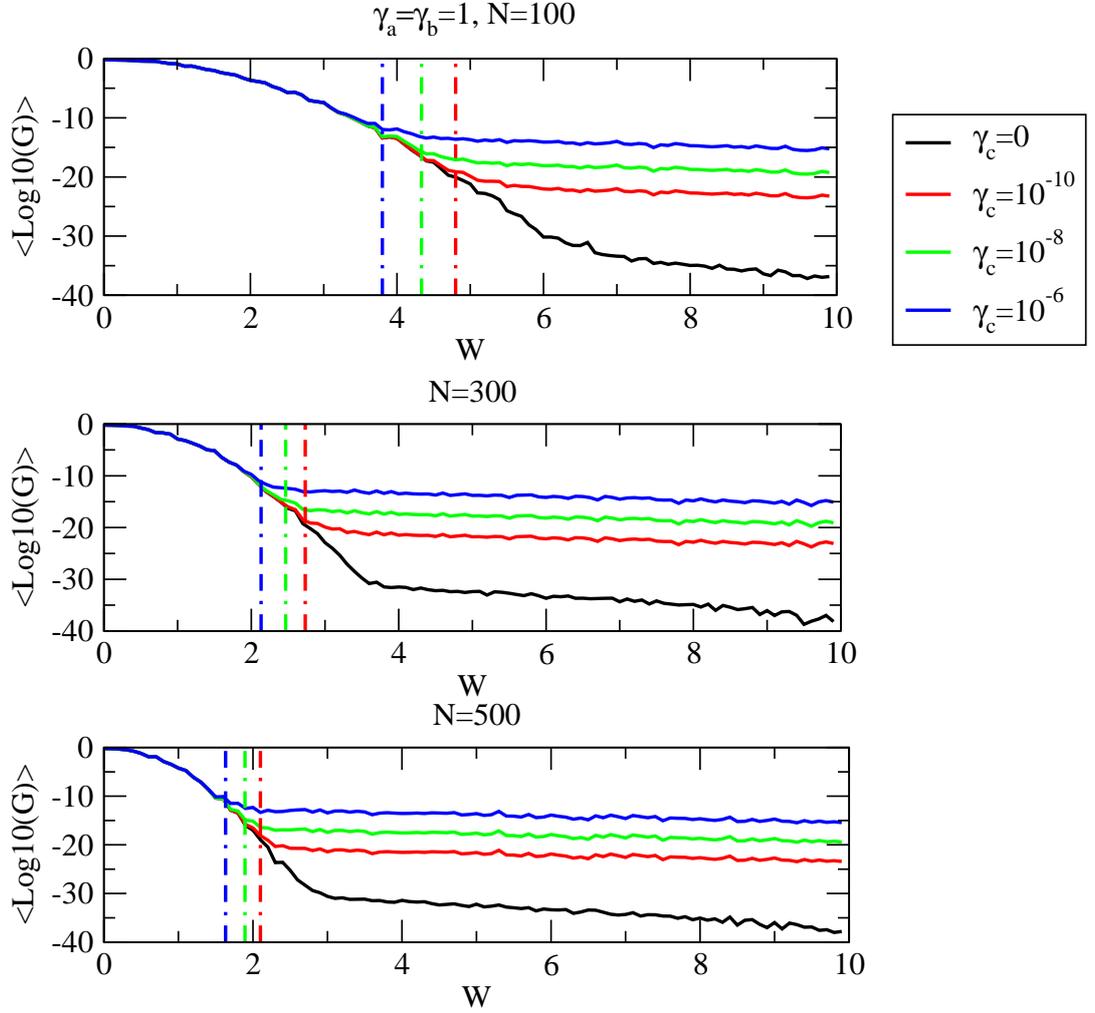}
\caption{$\braket{\log G (E=0)}$ vs $W$ for different
values of the coupling strength to common decay channel $\gamma_c$ and for different values of $N$ as indicated in the legend.
Here we average over $100$ disorder configurations for $N=100$ and over $50$ configurations
for $N=300$ and $N=500$. 
The coupling strength with edge channels is set $\gamma_a=\gamma_b=1$.
The dashed vertical line represents the critical value $W_c$, provided by the criterium
(\ref{crit}), for which the system becomes sensitive to opening, i.e. the behavior is different from the closed Anderson model. The value provided by the above criterium have to be rescaled by the factor $3/2$.
}
\label{fig:criticalw}
\end{figure}
In Figure \ref{fig:criticalw} we have computed the behavior of $G$ in function 
of the disorder strength $W$ in the limit $\kappa\ll1$.

We can see that if the system is not coupled to the common decay channel the conductance decreases as 
the the disorder increases. That makes sense because as $W$ is increased 
the eigenstates became more and more localized and so the transport 
from channel $a$ to channel $b$ is inhibited.
Otherwise if we switch on the coupling $\gamma_c$ we note that 
a critical amount of disorder $W_c$ exists, such that, for $W<W_c$, $G$
behaves in the same way as for a closed system.

From the expression of the eigenstates of $\he$ for small $\kappa$, see Eq. (\ref{pert1}), we can in principle estimate the value of $N$, $W$ or $\gamma$ for which the effect of the long range coupling becomes important. Indeed, as we increase $N$ or $W$, the wave function becomes localized,
$|\psi(j)| \sim \exp(-|j-j_0|/\xi)$. 
The probability to be at the edge sites for a state localized in the middle of the chain goes like $\exp(-N/\xi)$, with $\xi \sim 105.2/W^2$. On the other side from the Eq.(\ref{pert1}) we have that for the coupling to the common channel the probability to find a state in one site is proportional to $(\gamma/W)^2$. So we can set the criterion:
\be
\label{crit}
\exp(-N/\xi)\approx\alpha \ \left(\frac{\gamma_c}{W}\right)^2,
\ee
where $\alpha$ is the proportionality constant.
From Eq.(\ref{crit}) we are able to deduce $W_c$.
This rough criterium provide us a value of $W_c$ which is in good agreement
with the numerical results for different value of $N$ and $\gamma_c$ as shown in Figure \ref{fig:criticalw}. 

%%%%%%%%%%%%%%%%%%%%%%%%%%%%%%%%%

\section{Conclusions}
\label{sec-conc}
We have studied the $1D$ Anderson model with all sites
coupled to a common decay channel (coherent dissipation). 
Our main motivation  was to understand the interplay of opening 
and disorder. 
  Increasing the disorder tends to localize the states. Increasing the opening, on the other hand, reduces the degree of localization, and in particular induces a superradiance transition, with the formation of a subradiant subspace and a superradiant state completely delocalized over all sites.
%As we increase the opening, a superradiant
%transition occurs, with the formation of a superradiant and a subradiant subspace
%While as we increase disorder, the states tend to localize,
%the effect of opening goes in the opposite direction,
 %inducing a superradiant state completely delocalized over all the sites. 
Our results show that, while for
small opening all the states tend to be similarly affected by the 
disorder, for large opening the superradiant state remains delocalized
even as the disorder increases, while the subradiant states
are much more affected by disorder, becoming more
localized as the disorder increases. We have explained these effects qualitatively,
mainly guided by  perturbation theory. Indeed we have shown that 
the superradiant state is not affected by disorder, up to a critical disorder 
strenght for which the superradiance effect disappears.
This is because its energy
is very distant, in the complex plane,  from the energies of
the subradiant states. 

There are different experiments \cite{aspect, bec} about the localization of matter waves or light waves which can confirm the findings presented in 
this chapter.

Another important reason that drives us to study more this
model is that, as we will show in the next chapter, 
we have found the same interesting features in the three-dimensional
Anderson model with coherent dissipation and also in a
cold atoms system \cite{kaiser}.
This means that the $1D$ system is  a really paradigmatic  model 
and a powerful tool in order to understand the interplay of superradiance 
and disorde.

%%%%%%%%%%%%%%

\chapter{Interplay of superradiance and disorder in the $3D$ Anderson model}
\label{chap:3d}

In analogy with the one-dimensional case we want to study
a $3D$ Anderson model in
presence of coherent dissipation, i.e. when the particle can escape from the
system at any site of the cubic lattice.

As in the case considered in Chapter \ref{chap:1d}, we have only one decay channel and all the couplings strength 
are equal. The effective Hamiltonian is given by
\be
\he=H_0 -\rmi\frac{\gamma}{2}Q,
\ee
Here, $H_0$, is the standard Anderson Hamiltonian similar to (\ref{and1})
and $Q_{ij}=1 \ \forall i,j$.
The eigenvalues of $\he$ are labelled as follow
\be
\label{3deigen}
\E_r=E_r -\frac\rmi2\Gamma_r.
\ee 

Compared with the closed system
the main difference with respect to the $1D$ case is that
here we have a critical value of the disorder strength for which
all the eigenstates are exponentially localized, see chapter \ref{chap:anderson}.
For a tight binding model with diagonal uncorrelated disorder we have
\cite{zdetsis}
\be
\frac{W_c}\Omega\approx16.5.
\ee

Let us analyze the interplay of superradiance and disorder, 
similar to what it has been done for the $1D$ case, see Chapter \ref{chap:1d}.

%\begin{itemize}
%\item{as the coupling between the discrete states and the continuum 
%increases, one observe a segrgation of the widths, i.e. ST takes place.
%Also in this case, for large coupling, we have only one superradiant mode with  $\Gamma\approx N\gamma$ and $N-1$ subradiant states with $\Gamma\approx 0$.  }
%\item{Superradiant and subradiant subspaces are affected in a different manner by disorder.
%Subradiant states are sensitive to Anderson localization while superradiant sate remain delocalized up to a critical disorder strenght for which the superradiance effect disappears.}
%\item{The structure of the averaged probability 
%distribution (APD) (defined in section \ref{sec:structure}) for large disorder in the limit of
%small opening
%($\kappa\ll1$) and large opening
%($\kappa\gg1$) are qualitative the same explained in section \ref{sec:structure}.
%}
%\end{itemize} 

\section{Transition to superradiance}
First of all let us show that, in this system, ST takes places after 
a certain value of coupling to the common decay channel.

\begin{figure}
\centering
\includegraphics[width=0.7\textwidth]{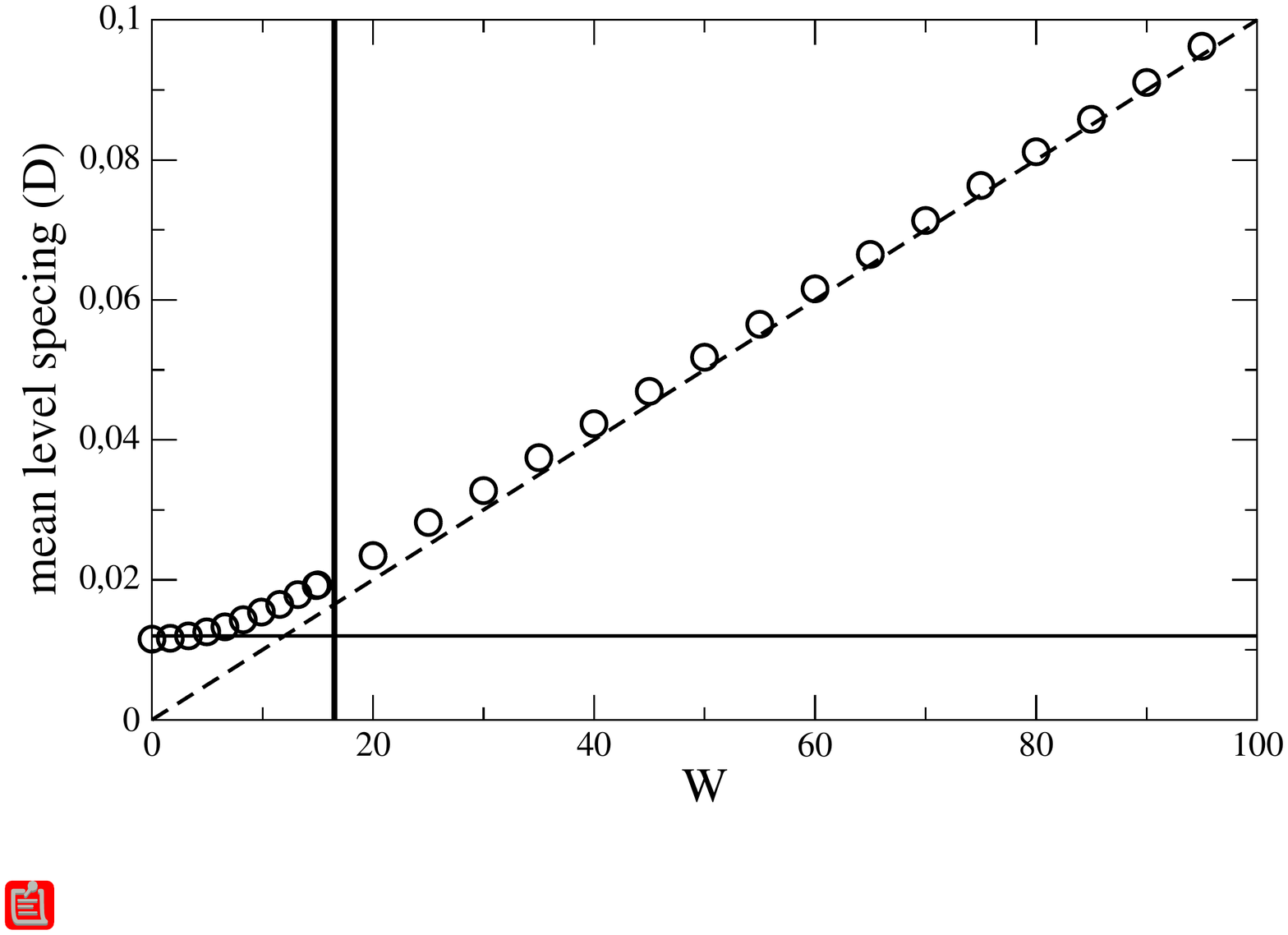}
\caption{The mean level spacing, $D$, as a function of $W$
for the closed system.
Here $N=10^3$, $\Omega=1$ and each point is obtained averaging over $10$ realizations. 
The horizontal solid line is $D=12\Omega/N$, the dashed line is $D=W/N$ and the vertical solid line is the Anderson transition ($W=16.5$). }
\label{fig:D}
\end{figure}

The ST is expected to occur for $\braket{\Gamma}/D\approx1$.
The average width is $\gamma$ and so
\be
\label{3dkappa2}
\kappa=\gamma/D, 
\ee
is the effective coupling strength to the continuum.
The mean level spacing as a function of the disorder strength is 
shown in Figure \ref{fig:D}.

For small disorder the mean level spacing is given by $D\approx 12\Omega/N$ while for large disorder we have $D\approx W/N$ 
and then
\be
\label{3dkappa}
\kappa\approx\frac{\gamma N}{W}.
\ee

\begin{figure}
\centering
\includegraphics[width=0.7\textwidth]{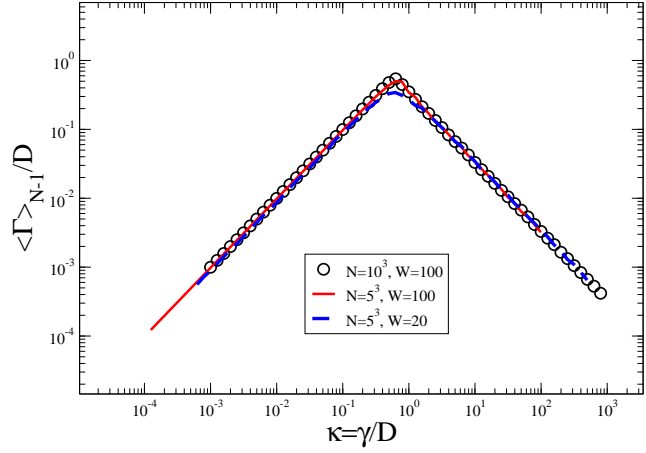}
\caption{The average width of the $N-1$ subradiant states, normalized by the mean level spacing $D$, versus the effective coupling strength $\kappa$ for different values of $N$ and $W$, and $\Omega = 1$. Here we averaged over $10$ disordered configurations.}
\label{fig:st}
\end{figure} 
In Fig. \ref{fig:st} we show that ST occurs at $\kappa\simeq1$ for different values of $W/\Omega$ and $N$.

\section{Sensitivity to disorder}
\begin{figure}
\centering
\includegraphics[width=01.2\textwidth]{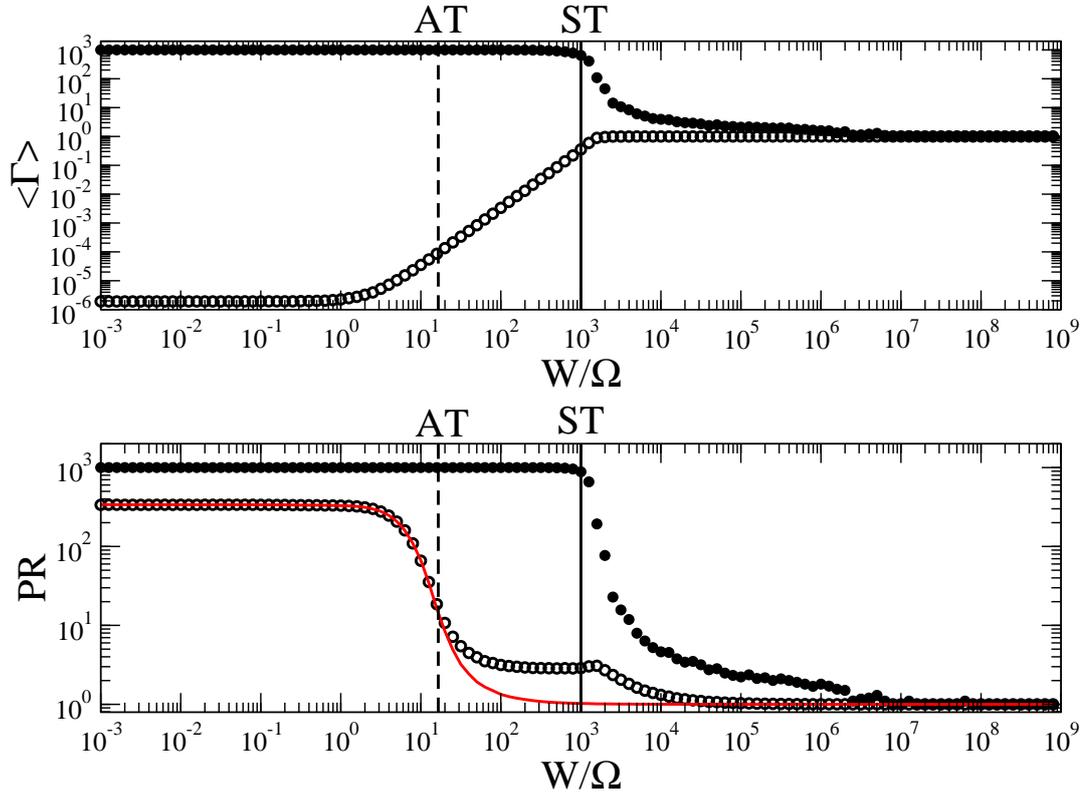}
\caption{Upper panel: the average width versus $W/\Omega$.
Lower panel: the participation ratio is shown as a function of the disorder strength $W/\Omega$. 
In both cases open circles stand for the subradiant states, full circles indicate the superradiant state, while the red line stand for the closed system. 
Each point is obtained by averaging over $10$ disorder realizations for the superradiant state, while for the subradiant states, an additional average over all the subradiant states is performed. For the closed system we have averaged over all the states. The right and left vertical lines indicate the superradiance transition (ST) and the Anderson transition (AT), respectively. Here the system was a cube of $10\times10\times10$ sites ($N=10^3$) and $\gamma = \Omega = 1$.}
\label{fig:3dpaper2}
\end{figure}

In the paradigmatic $1D$ model we have found that subradiant and superradiant subspace are affected 
by the disorder in a different way. Subradiant states are sensitive to Anderson localization, like the eigenstates of the closed system, while the superradiant state is unaffected up to the superradiance transition.
In order to study whether such a behavior occurs also in the $3D$ Anderson model we have computed the 
$PR$, defined in Eq.(\ref{pr}), as a function of disorder.
The numerical results are shown in Figure \ref{fig:3dpaper2}.

In order to show how the disorder affects the superradiant and the subradinat subspaces, 
in Figure \ref{fig:3dpaper2} we analyze both the participation ratio of the eigenstates of the effective Hamiltonian (PR), see Eq. (\ref{pr}), and the average decay widths, see Eq. (\ref{avewidth}).

Since in our model there is only one decay channel we have only one
superradiant state and $N-1$ subradiant states.

For each disorder strength we consider an average over $10$ realizations for the superradiant state, while for the subradiant states, an additional average over all the subradiant states is performed.

We have choosen $\gamma$ such that for small disorder, $W\ll1$,
we are in deep superradiant regime, i.e. $\kappa=\gamma/D\gg1$.
Increasing $W$ we reach the Anderson transition threshold ($W\approx 16.5$), see the dashed vertical line in Figure \ref{fig:3dpaper2}.
Above this critical value all the eigenstates of the closed model 
are exponentially localized but we are still in the superradiant regime.

If we increase the disorder strength further the mean level spacing increases as we have shown in Figure \ref{fig:D},
and for large disorder the mean level spacing becomes proportional to $W$. 
Since the coupling to the continuum behaves as described by Eq.(\ref{3dkappa}),
for $W>\gamma N$ the superdiant effect disappears 
because $\kappa<1$.

Summarizing, increasing $W$ from $W\ll1$ to $W\gg1$, we span three 
different regimes labeled as: 
\emph{delocalized superradiant} regime for ($W<16.5$),    \emph{localized superradiant} regime (for  $16.5<W<\gamma N$),
 and  \emph{localized not superradiant} regime for ($W>\gamma N$). 
Note that for   $W<16.5$, the eigenstates of the closed system can be localized or not, depending on their energy eigenvalue, see discussion in Chap \ref{chap:anderson},
while for $W>16.5$ all the states of the closed system are localized.

In the lower panel the $PR$ is plotted as a function of $W$.
Here also the $PR$ of the closed system is shown, see the red line.
For the closed system we have averaged over all the eigenstates.
 
In the delocalized superradiant regime the superradiant state is fully 
delocalized ($PR=N$) and the subradiant states are extended as the 
eigenstates of the closed model.

If we increase $W$ we cross the Anderson transition threshold and 
we enter in the localized superradiant regime.
The $PR$ of the subradiants states starts to decrease while the superradiant is not affected by this transition and it remains completely 
delocalized. 

If $W$ is increased further we enter in the localized not superradiant regime.
Only above this critical value the superradiant state starts to localize and, only for very large disorder (corresponding to very small 
$\kappa$), they behave in the same way as the subradiant states.

We also note that the subradiant states are affected by disorder as the eigenstates of the closed
system in the delocalized superradiant regime, i.e. for $W<16.5$.
This behavior can be viewed as a signature of the fact that the subradiant states are effectively decoupled
from the external world and therefore similar to the states of the closed system.

In the upper panel we show the average decay width of the 
subradiant  and superradiant states.
In the localized superradiant regime the average width of the superradiant states is approximately $\gamma N$, while the average width of the subradiants states is very small.
Here the average width of the two subspaces are well separated, signalling that we are in the superradiant regime.

When we enter in the localized superradiant regime the average width 
of the subradiant states starts to increase 
reaching the constant value $\gamma$ while the average width of
the superradiant state remains basically the same.
Also in this regime the average width of the two subspaces are segregate but we can see that the disorder tends to destroy the 
subradiant states.
Finally, when the disorder is such that $\kappa<1$, we enter in the
localized not superradiant regime and also the superradiant state starts to decrease its width.
Above this critical disorder strength the average width of the two subspaces became equal reaching the common value $\gamma$.

The regime for which the sensibility to disorder of the superradiant state and
the subradiant states  strongly differs is the
\emph{localized superradiant} regime. It is interesting to note that as we increase $N$, 
while the value of $W$ for which Anderson transition takes place, does not change,
the value of $W$ for which the ST takes place, increases proportionally to $\gamma N$.
This means that the range of $W$ where the 
two subspaces have different sensibility to disorder can be increased on increasing $N$ or $\gamma$.

In conclusion: in the closed $3D$ Anderson model, the localization properties of 
the system are different from $1D$ case because
all states became localized above a critical value {\bf independent of $\mathbf{N}$}.
Nevertheless the sensitivity to disorder of the two subspaces of the open model is exactly
the same: subradiant states are sensitive to Anderson localization, like the eigenstates of the closed system, while the superradiant state is sensitive only to the ST.

\begin{figure}[H]
\centering
\subfigure[]
{\includegraphics[width=10cm, height=6cm]{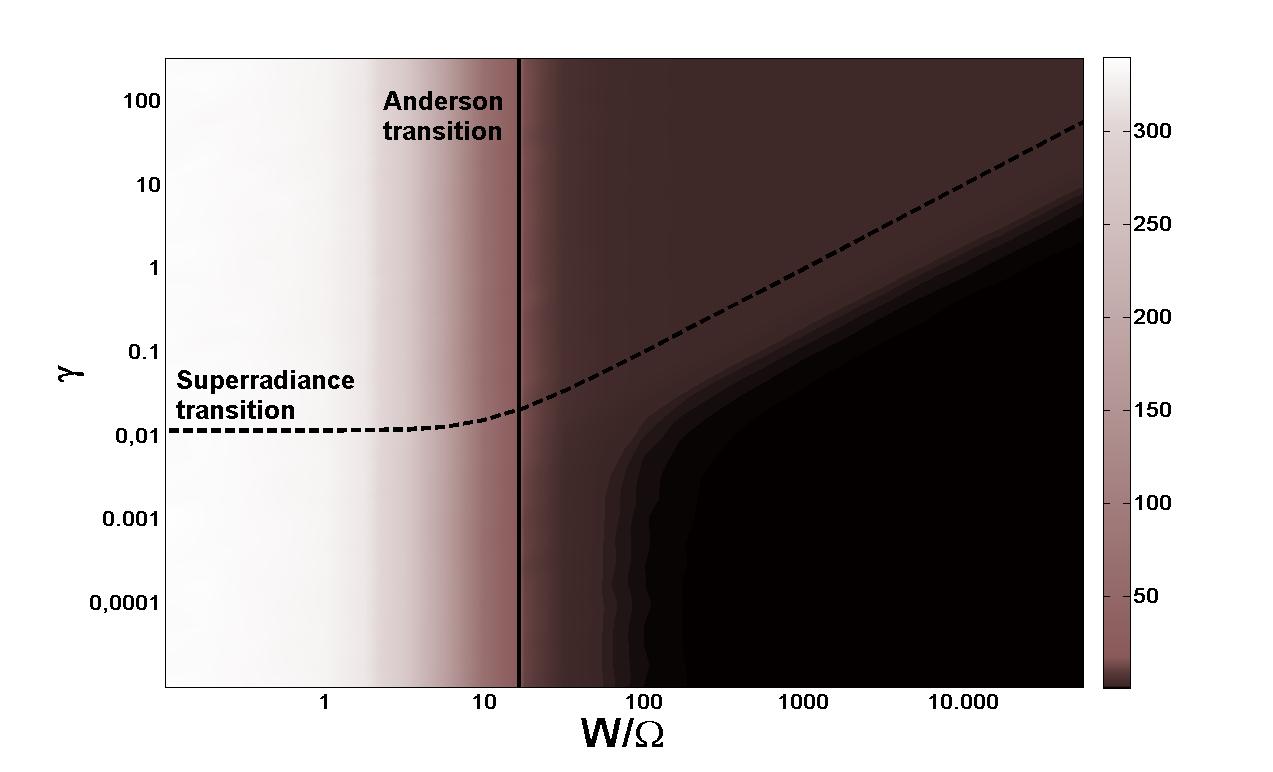}}
\hspace{1mm}
\subfigure[]
{\includegraphics[width=10cm, height=6cm]{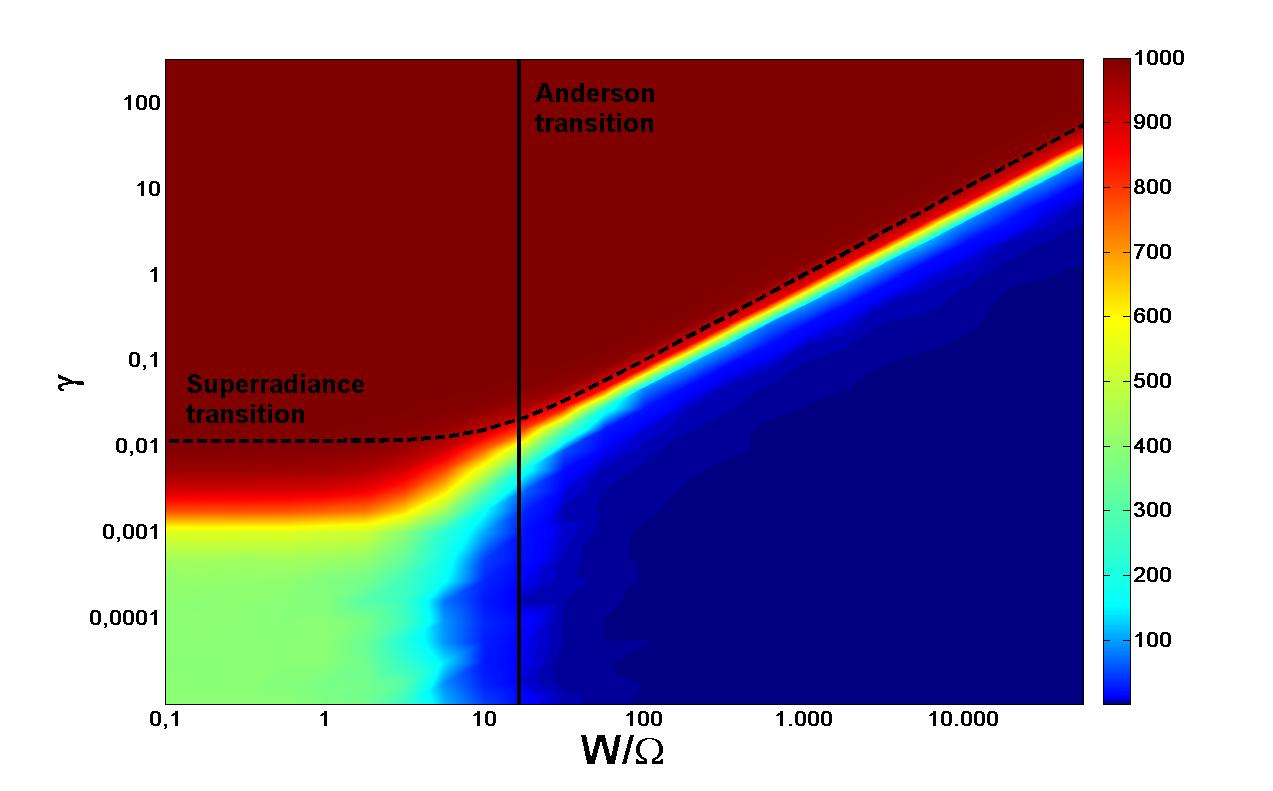}}
\caption{(a) PR of subradiant states as function of $\gamma$ and $W/\Omega$;
(b) PR of superradiant states as function of $\gamma$ and $W/\Omega$.
The dashed vertical line is the Anderson transition ($W=16.5$) and the solid line is the superradince transition ($\gamma=D$).
In both cases the system was a cube of $10\times10\times10$ sites ($N=10^3$) and each data is obtained averaging over $5$ realizations for the superradiant state while for the subradiant states an additional average over all the subradiant subspace has been done. }
\label{fig:phases3}
\end{figure}

In this section we have focused our attention on the degree 
of localization of the eigenstates of $\he$ mainly using the PR.
In Figure \ref{fig:phases3} our results are summarized:
the PR is shown as a function
of both the opening $\gamma$ and the disorder strength $W/\Omega$
for a fixed size of the system.

Note that we use two different colors scale for the two panels in 
order to increase the visibility.

Is important to note that what was known is only the behavior
of the PR for the closed Anderson model, i.e. 
on the $W/\Omega$ axis while
here we have fully analyzed the whole $(W/\Omega, \gamma)$ plane.

\section{Structure of wave functions}
For the $1D$ Anderson with coherent dissipation model we have found that 
in the limit of large disorder the structure of the 
averaged probability distribution (APD) is the following:
\begin{itemize}
\item{for small opening ($\kappa\ll1$) it consists in a sum of the APD 
of the closed Anderson model  and a plateau proportional to $(\gamma/W)^2$, the latter being independent of $N$.}
\item{for large opening ($\kappa\gg1$) it consists in a sum of the APD of the closed Anderson model and a plateau
proportional to $1/N$ and independent of $\gamma$ and $W$.}
\end{itemize}

\begin{figure}
\centering
\includegraphics[width=1\textwidth]{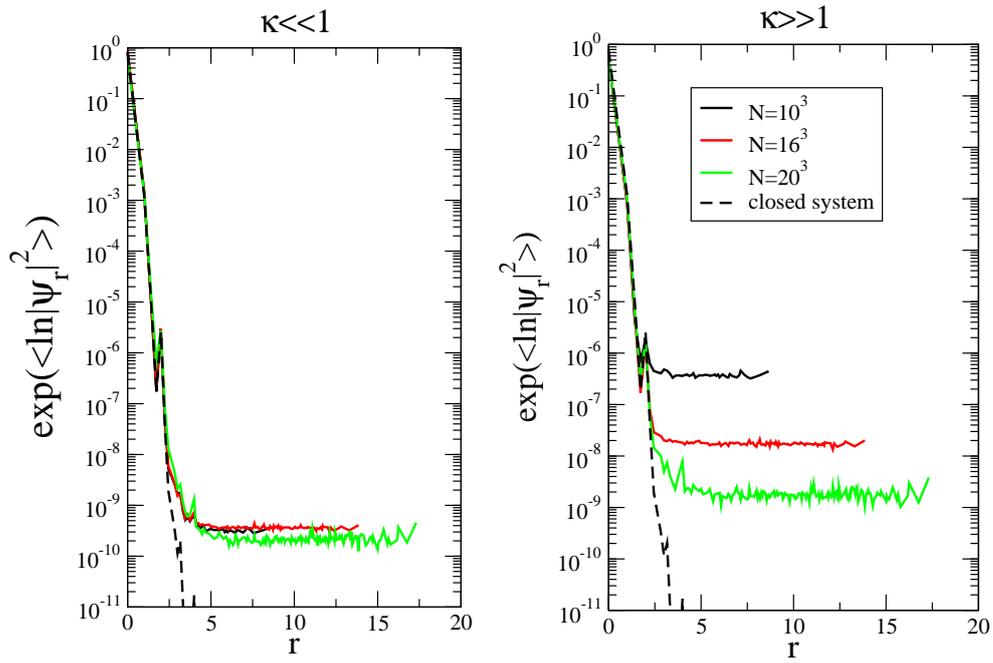}
\caption{APD of all the eigenstates of $\he$ that are strongly peaked in the center of the lattice as a function of the distance from the center of the lattice $r$. Here we fix $\gamma=10^{-3}$ in the left panel and $\gamma=10^2$ in the right panel. In both cases $W=100$. Here we averaged over $10$ spatial configurations. Moreover, to reduce fluctuations, we averaged the logarithm of the probability distribution.
}
\label{fig:openeigen}
\end{figure}

In order to study the typical APD of the eigenstates in the $3D$ system 
we have studied the APD of the eigenstates of $\he$ that are strongly
peaked in the center of the cubic lattice.
The APD of these states are plotted in Figure \ref{fig:openeigen} as a function of
the distance from the center of the lattice.
We can see from Figure \ref{fig:openeigen} that the APD of the 
eigenstates peaked in the middle of the lattice is very close to the APD 
of the one-dimensional model.

For small opening ($\kappa\ll1$) we have a plateau independent of 
the size of the cube while for large opening ($\kappa\gg1$) 
the plateau decrease as $N^{-1}$.
Even in this case the peak of the APD has exactly the shape of 
the closed model (see the dashed line in Figure \ref{fig:openeigen}) and it depends from the parameters $W$ and $E$
in the same way.

\begin{figure}
\centering
\includegraphics[width=1\textwidth]{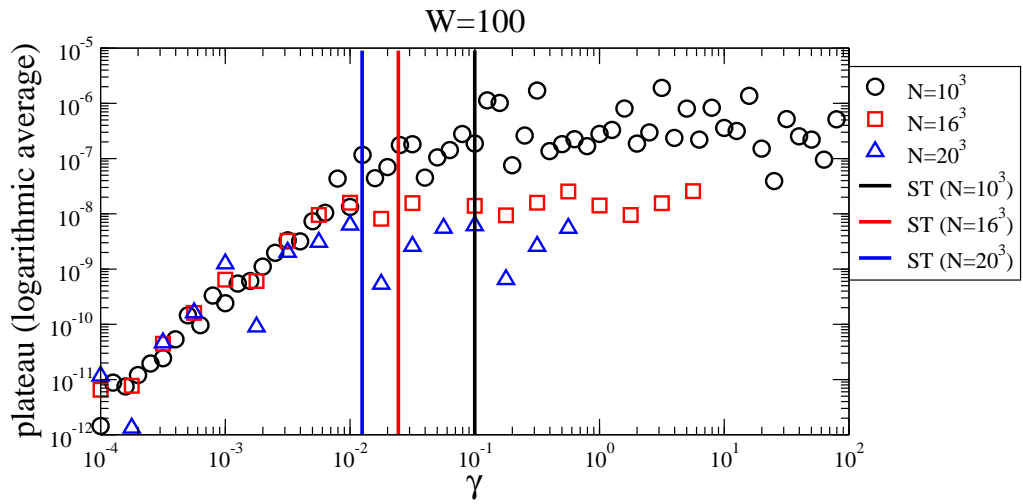}
\caption{The height of the plateau as a function of $\gamma$ for different $N$.
Here $W=100$ and each point is obtained averaging over 
$10$ disorder realizations for $N=10^3$ and $N=16^3$ and $5$ realizations for $N=20^3$.
The vertical solid line are the STs for different N, the colors reflect the different numerical data.
}
\label{fig:plateauw100}
\end{figure}

For these reasons the framework developed for the $1D$ case (see subsection \ref{sec:structure}) can be very useful for this system too.
In Figure \ref{fig:plateauw100} the plateau part of the APD is shown as a function of $\gamma$ for different $N$.

\chapter{Conclusions}
The degree of opening and intrinsic disorder strongly affects the proprieties of mesoscopic systems in the quantum coherent regime.
While in literature both opening and disorder have been extensively studied, their interplay has been not sufficientely addressed.

In this thesis we have studied their interplay
in the $1D$ and $3D$ Anderson model with coherent dissipation and we have found several results, most of them not yet published.
Some results about the $1D$ system have been published on the special issue of 
\emph{Progress of Physics} with the title\emph{
``Quantum Physics with Non-Hermitian Operators: Theory
and Experiment''}, see \cite{biella}.
\newline
\newline
\textbf{$1D$ Anderson model with coherent dissipation}
\newline
\newline 
While the disorder tends to localize the states, the opening reduces the degree of localization and in particular induces a superradiance transition, with the formation of a subradiant subspace and a superradiant state completely delocalized over all sites.

We have shown that, while for small opening all the eigenstates are
similarly affected by disorder, for large opening the
superradiant state and the
subradiant states have a very different sensitivity to disorder.
Subradiant states are sensitive to localization transition (as the eigenstates 
of the closed system) while the superradiant states remains delocalized until the
superradiance effect disappears. 

The regime for which the behavior of the two subspaces is  different is the \emph{localized superradiant} regime,
which is located between Anderson transition and the superradiance transition.

In this regime we have both superradiant and subradiant states
($\kappa>1$),
and the disorder strength is such that all the eigenstates of the closed system are exponentially localized with localization length 
smaller than the system size ($\xi<N$).

In the $1D$ Anderson model the localization threshold and the superradiance transition depends both on the sample size.
Indeed if the degree of the opening ($\gamma$) is fixed
the disorder strength for which $\xi<N$
behaves as
\be
W_c\propto\frac1{\sqrt{N}}
\ee
and the disorder strength for which $\kappa>1$ is proportional to $N$
\be
W_{\text{ST}}\propto N.
\ee

For this reason
increasing $N$ is possible to enlarge the range of the parameters
where we have the localized superradiant regime.

All these informations are well summarized in Figure \ref{fig:phases2}
where the participation ratio (PR) is shown as a function of
opening and disorder ($\gamma$ and $W/\Omega$).
We remark that the behavior of the PR vs $W$ for $\gamma=0$ was
known. Also the behavior of PR vs. $\gamma$ for $W=0$ was known.
Here we have fully analyzed the whole $(W,\gamma)$ plane.

We have also characterized the structure of the averaged probability distribution (APD) in the limit of large disorder:
\begin{itemize}
\item{
for small opening ($\kappa\ll1$ ) it consists of two terms: the APD of the closed 
Anderson model and a plateau proportional to $(\gamma/W)^2$ and independent of $N$.}
\item{
For large opening ($\kappa\gg1$) it consist of two terms: the APD of the closed Anderson model and a plateau proportional to $1/N$ and independent of $\gamma$ and $W$.}
\end{itemize}
The APD can be measured experimentally because it can be related to 
the density in the case of non-interacting particles.
\newline
\newline
\textbf{$3D$ Anderson model with coherent dissipation}
\newline
\newline
In this system we have shown that the general features found in the $1D$
model holds also in three dimensions.

The different sensitivity to disorder of the two subspaces is confirmed
and the structure of the averaged probability distribution is similar to the one-dimensional case.

Even in the $3D$ case, the regime for which the behavior of the two subspaces is  different is the \emph{localized superradiant} regime
namely the regime for which superradiance transition is reached 
and all the eigenstates of the closed system are localized.

The main difference is that in $3D$ Anderson model the disorder strength 
for which Anderson transition takes place ($W\approx16.5$) is independent of $N$.
Nevertheless if the degree of opening is fixed the superradiance threshold depends linearly of $N$ 
\be
W_{\text{ST}}\propto N.
\ee
This means that in $3D$ is possible to enlarge the range of the parameters
for which the localized superradiant regime is reached 
increasing the size of the cubic lattice.
\newline
\newline
\textbf{Apllications and perspectives}
\newline
\newline
The findings of our work can be used for instance to study the propagation of a photon through a cold atoms cloud:
if the wavelength of the photon is larger than the system size an effective non-Hermitian long-range interaction is created.
Concerning the cold atoms system, our preliminary results, confirm the general behaviors found in our \emph{symple} model. 
This is not trivial because the cold atoms are not decribed by Anderson model.
This confirm that our model is paradigmatic and can be used to
understand the general features of more realistic systems.  

Another important application involves the photosynthetic complexes
where the transport proprieties are strongly affected by the interplay of superradiance and static disorder.

Our results can be applied in all those mesoscopic systems
(in the quantum coherent regime) where opening and disorder are considered and for which 
the wavelength of the particle is larger than the typical length scale of the system.

In perspectives we plan to study also the interplay of superradiance and dynamic disorder (dephasing) instead of static disorder with direct application for quantum computation and energy transport in light harvesting systems.

\begin{mychapter}[Ringraziamenti]
Queste pagine sono il risultato di qualche anno di passione e studio:
ovviamente sarebbe stato impossibile poter portare a termine questo \emph{viaggio} senza l'aiuto (consapevole o no) di persone, posti, viaggi, oggetti, pensieri, sogni e avvenimenti casuali. 
\newline
Per questo motivo tengo a ringraziare (in ordine sparso e omettendo posti, viaggi, oggetti, pensieri, sogni e avvenimenti casuali):
\newline

la mia famiglia, senza di loro sarebbe stato tutto molto più complicato, mi ha sempre sostenuto e reso libero
di fare le mie scelte, sempre.
Un particolare \emph{grazie} a mio padre:
mi piace pensare che sia stato lui a insegnarmi la \emph{passione}.
\newline

I miei amici, mille cose sono successe in questi cinque anni.
Grazie per tutte le notti passate insieme,
per tutte le risate e per tutte le cose che non si possono scrivere nei ringraziamenti di una tesi.
\newline

La \emph{luna}: l'unico satellite naturale della Terra.
\newline

I miei compagni di viaggio del DMF, 
condividere lo stesso posto tutti i giorni, per cinque anni, ci ha reso 
una piccola famiglia. Le scadenze non potranno cambiare questo.
\newline

Ringrazio il Prof. Fausto Borgonovi  per tutti i concetti che mi ha insegnato ma soprattutto per avermi mostrato un metodo di lavoro.\newline

Ringrazio il Dott. Luca Celardo per tutte le ore che nell'ultimo anno mi ha dedicato e per la sua infinita disponibilità.
Mi sdebiterò in un futuro.
\newline

Infine, ringrazio me stesso, per aver tenuto botta, 
nonostante tutto.

\end{mychapter}

\backmatter

\end{document}